\documentclass{article}

\usepackage{graphicx}           % required for inserting images
\usepackage{geometry}           % decrease page margins
\usepackage{authblk}            % authors and affiliations
\usepackage[numbers]{natbib}    % references
\usepackage{amsmath}            % math
\usepackage{amssymb}            % additional math symbols
\usepackage{bbm}                % more math symbols
\usepackage{graphicx}           % figures
\usepackage{multirow}           % multirow tables
\usepackage{mhchem}             % chemical equations
\usepackage{tikz}               % figures
\usepackage{algorithm}          % algorithms
\usepackage{algpseudocode}
\usepackage{booktabs}

\title{Safe Deployment of Offline Reinforcement Learning via Input Convex Action Correction}
\author[1]{Alex Durkin}
\author[2]{Jasper Stolte}
\author[2]{Matthew Jones}
\author[3]{Raghuraman Pitchumani}
\author[2]{Bei Li}
\author[2]{Christian Michler}
\author[1]{Mehmet Mercang\"{o}z}
\affil[1]{Department of Chemical Engineering, Imperial College London, SW7 2AZ, UK}
\affil[2]{Shell Information Technology International BV, 1031 HW Amsterdam, NL}
\affil[3]{Shell Global Solutions International BV, 1031 HW Amsterdam, NL}
\date{}

\begin{document}

\maketitle

\section*{Abstract}

Offline reinforcement learning (offline RL) offers a promising framework for developing control strategies in chemical process systems using historical data, without the risks or costs of online experimentation. This work investigates the application of offline RL to the safe and efficient control of an exothermic polymerisation continuous stirred-tank reactor. We introduce a Gymnasium-compatible simulation environment that captures the reactor's nonlinear dynamics, including reaction kinetics, energy balances, and operational constraints. The environment supports three industrially relevant scenarios: startup, grade change down, and grade change up. It also includes reproducible offline datasets generated from proportional-integral controllers with randomised tunings, providing a benchmark for evaluating offline RL algorithms in realistic process control tasks.

We assess behaviour cloning and implicit Q-learning as baseline algorithms, highlighting the challenges offline agents face, including steady-state offsets and degraded performance near setpoints. To address these issues, we propose a novel deployment-time safety layer that performs gradient-based action correction using input convex neural networks (PICNNs) as learned cost models. The PICNN enables real-time, differentiable correction of policy actions by descending a convex, state-conditioned cost surface, without requiring retraining or environment interaction.

Experimental results show that offline RL, particularly when combined with convex action correction, can outperform traditional control approaches and maintain stability across all scenarios. These findings demonstrate the feasibility of integrating offline RL with interpretable and safety-aware corrections for high-stakes chemical process control, and lay the groundwork for more reliable data-driven automation in industrial systems.

\subsection*{Keywords}
Offline Reinforcement Learning, Process Control, Safety, Convex Neural Networks

\section{Introduction}

The application of reinforcement learning (RL, \cite{sutton2018reinforcement}) to industrial process control holds promise for improving efficiency, adaptability, and long-term performance. However, real-world deployment of RL policies in such settings remains rare, particularly in high-stakes domains like chemical processing, power generation, and autonomous manufacturing, due to strict safety and stability requirements \cite{baldea2025automated}. This concern is heightened in the context of offline RL, where policies are trained entirely from historical data with no opportunity for online trial-and-error \cite{amodei2016concrete}. While this mitigates the risk of unsafe exploration during training, it also introduces the challenge of verifying and correcting policy behaviour at deployment time, when unseen disturbances or distributional shifts can lead to failure.

In industrial systems, a poorly chosen control action can have serious consequences, including damaging equipment, violating regulatory constraints, or causing cascading system instability. Offline-trained RL agents, optimised for historical performance, may produce unsafe or unstable actions when operating in regions underrepresented in the training data. Therefore, ensuring safety at deployment becomes a critical bottleneck in applying RL to real-world process control.

Polymerisation grade transitions are critical but challenging operations in continuous reactors, requiring changes in feed composition, temperature, and other conditions to switch from one polymer product to another \cite{bonvin2005optimal}. These transitions are nonlinear, constrained, and economically sensitive: off-specification material generated during the switch is wasteful, while long transition times reduce throughput \cite{lee2003iterative, prata2008integrated}. Managing grade changes efficiently demands controllers that balance safety, product quality, and operational speed—making them an ideal testbed for intelligent control methods such as RL.

Control theory offers powerful tools to address this challenge. In particular, the concepts of asymptotic stability and regions of attraction are foundational in guaranteeing that a control policy not only keeps the system bounded but also drives it to a desired operating regime \cite{lyapunov1992general}. These properties are often certified using Lyapunov functions, which provide a scalar energy-like measure that decreases over time under the system dynamics, ensuring stability and safe recoverability from perturbations \cite{khalil1996nonlinear}.

Inspired by this perspective, our approach introduces a deployment-time safety mechanism for offline RL agents using partially input convex neural networks (PICNNs, \cite{amos2017input}) as learned cost models. These models are constructed to be convex in the control actions and expressive over the state, enabling efficient, gradient-based correction of unsafe actions. When the offline RL policy proposes an action that may compromise stability or violate safety constraints, the PICNN model provides a locally optimal, safe correction by descending its convex cost surface. Although not explicitly trained as Lyapunov functions, the structure of these models ensures they behave like a learned control energy landscape that discourages unsafe actions and guides the system toward safe operating regions.

This architecture is particularly well-suited for process control applications, where control inputs must be interpretable, verifiable, and responsive to safety constraints. Our method acts as a safety layer over arbitrary offline policies, enabling stable and conservative control corrections in real time without requiring modifications to the underlying policy or additional interaction with the environment.

This paper proceeds with a summary of the related work in literature followed by a summary of our contributions. Section \ref{sec:background} provides some background on offline RL and PICNNs for safe control. Section \ref{sec:benchmark} describes the benchmark environment and datasets used for evaluation. Section \ref{sec:offline} details the offline training of RL agents, including behaviour cloning and implicit Q-learning as well as the PICNN cost model. Section \ref{sec:online} outlines our novel online action correction algorithm. Section \ref{sec:results} presents the experimental results, comparing the performance of offline RL agents with and without PICNN-based action correction. Finally, Section \ref{sec:conclusion} concludes with a discussion of the findings and future work.

\subsection{Related Work}

A range of methods has been proposed to incorporate safety into RL, especially in the context of continuous control. A prominent direction is the use of Lyapunov-based techniques to constrain policy updates or verify system stability. For example, \cite{berkenkamp2017safe} introduced a model-based RL framework that leverages Lyapunov functions to guarantee safe learning. Their method models uncertainty in the dynamics using Gaussian processes to estimate dynamics and enforce stability through Lyapunov constraints. Other efforts apply control barrier functions (CBFs) to ensure constraint satisfaction through quadratic programming \cite{ames2017control}, and some integrate CBFs with control Lyapunov functions to achieve safety and stability simultaneously \cite{gangopadhyay2023safe}. These techniques offer strong guarantees but often assume access to accurate dynamics models and are primarily applied during training or online control.

Another branch of safe RL research focuses on constrained policy optimisation. Constrained policy optimisation \cite{achiam2017constrained} ensures monotonic policy improvement while respecting safety constraints in expectation. Other works adopt primal-dual or Lagrangian methods to dynamically balance reward maximisation and constraint enforcement \cite{dalal2018safe}. In the offline RL setting, safety becomes even more critical due to the absence of exploration. Conservative Q-learning \cite{kumar2020conservative}, implicit Q-learning \cite{kostrikov2021offline}, and model-based offline policy optimisation \cite{yu2020mopo} address distributional shift and unsafe extrapolation by constraining policies to remain close to the behaviour policy or trusted regions of the dataset. While these methods reduce the risk of catastrophic actions during inference, they lack mechanisms to actively correct unsafe outputs at deployment.

Other approaches learn explicit safety critics or constraint models. For example, Lyapunov-based actor-critic methods \cite{wang2024actor} train neural approximations of Lyapunov functions to evaluate and constrain policy updates. Risk-sensitive methods model quantiles or variance of returns to encode safety preferences \cite{dabney2017distributional}. Action correction techniques such as those in \cite{dalal2018safe} modify unsafe actions using learned penalty gradients, while adaptive shielding mechanisms preemptively reject or reroute unsafe behaviours using external models \cite{liu2025ekg}. These methods provide a degree of runtime robustness but often require constraint supervision or multiple value functions, increasing complexity.

To improve action selection reliability, several works introduce architectural biases into RL function approximators. It has been demonstrated that PICNNs can be used in the context of RL to model the negative action-value function, as a convex function over the action space \cite{amos2017input}. This enables efficient inference of optimal actions through convex optimisation, making the architecture particularly suitable for structured decision-making tasks. The convex structure ensures global optima and promotes stable action selection, a desirable property for safe control. Follow-up work has extended PICNNs to learn convex control models \cite{chen2019optimal} or embed differentiable convex solvers into policy networks \cite{agrawal2019differentiable}. Related architectures like OptNet \cite{amos2021optnet} and differentiable quadratic programming layers \cite{agrawal2019differentiable} allow NNs to output solutions to embedded convex programs, opening new paths for safe and structured control.

Finally, applying RL to real-world process control systems introduces unique demands that go beyond typical benchmark settings. These systems are highly sensitive to even brief periods of instability, and safety constraints are often implicit, non-differentiable, or time-varying. In industrial practice, model predictive control (MPC) is the prevailing strategy for handling such complexities. For example, a polyethylene producer reported reducing transition times by 25--50 \%, cutting off-spec product, and increasing throughput by 7 \% using an MPC system combined with real-time optimisation \cite{bonvin2005optimal}. Hybrid approaches that fuse mechanistic models with data-driven components, such as NNs augmenting mass and energy balances, have also been deployed to provide virtual sensors and improve predictive accuracy during polymer grade transitions \cite{benamor2004polymer}. More recently, RL agents trained in simulation have shown performance on par with tuned nonlinear MPC controllers for managing grade changes, highlighting their potential for data-driven process optimisation \cite{jiang2023polymer}.

Several benchmark environments have been developed to support safe and realistic RL research in these domains. The Industrial Benchmark \cite{hein2017benchmark}, Safety Gym \cite{ray2019benchmarking}, and PC-Gym \cite{bloor2024pc} all simulate control tasks with realistic dynamics, constraints, and disturbances. Prior work emphasises the importance of deployment-time guarantees, real-time corrective mechanisms, and interpretable architectures to bridge the gap between simulation and operational deployment \cite{cheng2019end}.

\subsection{Our Contributions}

Building on prior work in Lyapunov-based safe control and convex neural architectures, we introduce a novel deployment-time safety mechanism tailored for offline RL policies in industrial process control environments. Rather than relying on handcrafted Lyapunov functions or using PICNNs to approximate Q-values during training, we reinterpret PICNNs as state-conditioned cost surfaces that guide real-time correction of unsafe actions.

Our approach constructs a convex energy landscape over the action space using a PICNN, which is expressive in state and convex in action. At deployment, when an offline policy proposes a control input, the PICNN provides a corrective gradient that descends the learned cost surface, nudging the action toward safer, more stable alternatives. This gradient-based filtering mechanism avoids the need for explicit constraint sets or online exploration and can be applied to any offline policy without retraining or modifying its structure.

To evaluate this approach, we develop a Gymnasium-compatible benchmark environment that models a continuous stirred-tank reactor undergoing exothermic polymerisation. The environment captures the nonlinear dynamics, energy balances, and safety-relevant operating constraints typical of industrial grade transition scenarios, providing a realistic testbed for data-driven control strategies. Offline datasets are generated using PI controllers with diverse tuning parameters to simulate suboptimal operational behaviour.

To our knowledge, this is the first method to use PICNNs as learned safety surrogates for deployment-time action correction in high-stakes control domains. Our system functions as a lightweight safety layer that is interpretable, responsive, and compatible with real-time constraints, offering a practical pathway to bridge the gap between offline RL and safety-critical real-world deployment.

\section{Background} \label{sec:background}
RL is the field within machine learning in which an agent learns to optimise its behaviour in an environment through interaction \cite{sutton2018reinforcement}. The key advantage of RL is that it can learn a control policy which deals with the trade off between greediness and patience to chase an even greater expected reward in the future. The theoretical framework underpinning RL is the Markov decision process (MDP).

\subsection{Markov decision process}
The RL problem can be formalised in the context of determining a policy $\pi(a | s)$ to maximise the cummulative reward of a MDP given by $\left< \mathcal{S}, \mathcal{A}, p_0(s), p(s' | s, a), r(s,a), \gamma \right>$, where $\mathcal{S}$ is the state space, $\mathcal{A}$ is the action space, $p_0(s)$ is the initial state distribution, $p(s' | s, a)$ is the transition function, $r(s, a)$ is the reward function, and $\gamma$ is the discount factor where more importance is placed on future rewards as $\gamma\rightarrow1$.
A common strategy to solve the MDP involves approximating the state--action value function, $Q(s, a)$, referred to as the Q-function, which is defined as the expected cumulative reward of taking action $a$ in state $s$ and following the policy $\pi$. The optimal policy, $\pi^*$, is then given by:

\begin{equation}
    \pi^*(a | s) = \arg \max_{a} Q(s, a)
\end{equation}

\begin{figure}[htb]
	\centering
	\includegraphics[width=0.7\textwidth]{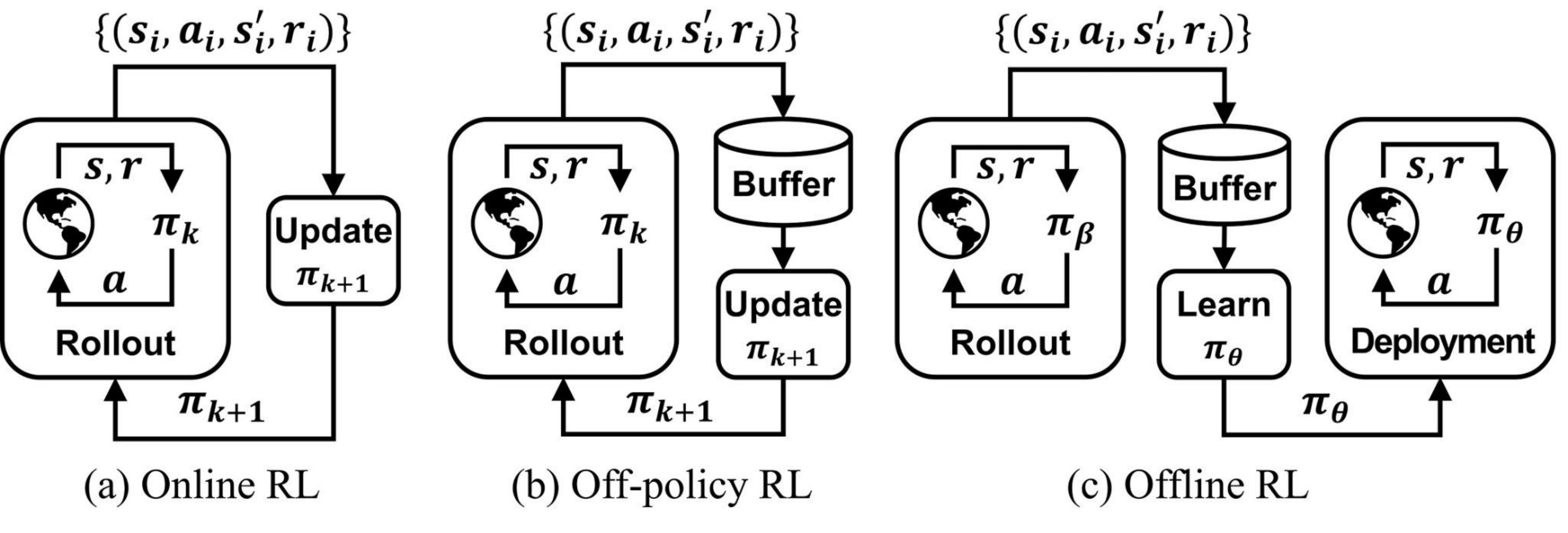}
	\caption{Three different classes of reinforcement learning (taken from \cite{kim2024offline}). Offline reinforcement learning is a special case of off-policy learning leveraging historic data only.}
	\label{fig:offline_vs_online}
\end{figure}

Process control applications involve continuous state--action spaces, with the value function control policies commonly modeled using NNs. Three distinct classifications of RL algorithms can be identified as demonstrated in figure \ref{fig:offline_vs_online}. In the usual case the agent can iterate towards a solution, collecting feedback from every interaction cycle. In off-policy algorithms a buffer of historic observations is preserved to increase sample efficiency and prevent convergence to local optima. Critically, in off-policy algorithms the agent still collects feedback on its current policy through rollouts of suboptimal agents during training. In chemical process control, deployment of partially trained agents on real production plants is infeasible for safety and economic reasons. Thus, we can either train an agent on a simulated process to subsequently deploy it on the real plant, or train an agent on data taken from the real process.

Using a simulator to train agents has several drawbacks. The most important one being that performance will suffer from the sim-to-real gap as realistic behaviour inherent to chemical processes is difficult to capture in the simulator. High fidelity dynamic simulation models are typically expensive to develop and maintain meaning they will not be available for most processes. These models are also slow leading to high compute requirements to train a policy. For these reasons, the authors propose training on historic process data (offline RL) as the preferred approach.

\subsection{Offline RL}
In offline RL \cite{levine2020offline, prudencio2023survey} a buffer of historic observations is used to train a policy for the agent. In this setting the data was collected under an unknown behaviour policy $\pi_\beta$, which could be any combination of manual operation with supervisory control. The objective is to learn a safe and robust control policy from this data without further environment interaction.

The key challenge with offline RL is \emph{distributional shift} \cite{kumar2019stabilizing}, which can be conceptually understood as the difficulty to accurately estimate the value of state--action combinations that were not observed in the historic data. To illustrate, consider the temporal difference loss to minimise the Bellman error for the Q-function as given below:

\begin{equation} \label{eq:bellman}
\mathcal{L}(\theta) = \mathbb{E}_{\mathcal{D}} \left[ 
\left( r + \gamma \max_{a'} Q_{\hat{\theta}}(s', a') - Q_{\theta}(s, a) \right)^2 
\right]
\end{equation}

where $\mathcal{D}$ is the dataset of (state, action, next state) observations, $Q_{\theta}$ is the parameterised Q-function prediction for the current state--action pair, and $r + \gamma \max_{a'} Q_{\hat{\theta}}$ is the Bellman equation for calculating the target Q-value as the sum of the immediate reward and the discounted maximum expected future rewards.

The challenge with distributional shift is that the loss in equation \ref{eq:bellman} contains a term $Q_{\hat{\theta}}(s',a')$ which requires evaluating the Q-function at state--action combinations that are not present in the data. Even worse, this Q-function evaluation is maximised over all possible future actions leading to gross value overestimation, learning policies that exploit this non-existing value.

Typical strategies to stabilise learning as employed in online RL include double Q-learning \cite{vanhasselt2015deep} and using target networks \cite{mnih2015human} or simply using vast amounts of training data.
These strategies can only slightly help reduce the magnitude of the value overestimation but do not fundamentally solve the problem.
Since this issue was clearly articulated in \cite{kumar2019stabilizing} in 2019 the field of offline RL gathered momentum and many strategies have been devised to develop policies with more robustness towards distributional shift.
Algorithms such as conservative Q-learning \cite{kumar2020conservative}, IQL \cite{kostrikov2021offline}, and behaviour-regularised actor-critic \cite{wu2019behavior} mitigate distributional shift by constraining policy updates toward the behaviour policy or learning conservative value estimates.

\subsection{Behaviour cloning}
BC is a straightforward imitation learning approach in which a policy is trained to directly mimic the actions taken by an expert or behaviour policy using supervised learning. BC ignores the reward signal entirely and only attempts to mimic the actions from the dataset as closely as possible. Formally, the BC loss function is given by equation \ref{eq:BC}

\begin{equation} \label{eq:BC}
    \mathcal{L}_{\text{BC}}(\phi) = \mathbb{E}_{(s, a) \sim \mathcal{D}} \left[ -\log \pi_\phi(a | s) \right]
\end{equation}

where $\mathcal{D}$ is the offline dataset of state--action pairs, and $\phi$ denotes the parameters of the policy network. If the dataset has good coverage of the state space and contains high-quality examples, BC can be expected to give good results. In the face of limited and/or subobtimal data BC is likely to give poor performance. Regardless, it should always be considered as a baseline in offline RL problems.

\subsection{Implicit Q-learning}
Implicit Q-learning (IQL) \cite{kostrikov2021offline} is an offline RL approach that avoids the distributional shift problem by never querying the target Q-function for state--action combinations that are not in the dataset. This fully in-sample learning depends on an approximation of the Q-value of the next state--action combination by a state-value function. IQL leverages expectile regression to predict an upper expectile of the state-value function with respect to the action distribution:

\begin{equation}
    \mathcal{L}_V(\psi) = \mathbb{E}_{\left(s, a\right) \sim \mathcal{D}} \left[ \mathcal{L}_2^\tau \left( Q_{\hat{\theta}}(s, a) - V_{\psi}(s) \right) \right]
\end{equation}

where $\mathcal{L}_2^\tau(u) = |\tau - \mathbbm{1}(u<0)|u^2$ is the expectile loss function, which is a generalisation of the mean squared error loss which enables asymmetric weightings to favour larger expectiles. If $\tau$ is equal to $0.5$, the expectile regression is identical to the normal MSE. As $\tau$ is increased towards $1$ the value no longer learns the average value of a given state over the action distribution but approaches the maximum value supported by the actions in the data. This estimated value function is then used to backup into the Q-function update instead of $Q_{\hat{\theta}}(s',a')$:

\begin{equation}
    \mathcal{L}_Q(\theta) = \mathbb{E}_{\left(s, a, s'\right) \sim \mathcal{D}} \left[ \left( r(s,a) + \gamma V_{\psi}(s') - Q_{\theta}(s, a) \right)^2 \right]
\end{equation}

Once an estimated Q-value function is learned, the policy maximising the expected value is extracted using AWR \cite{peng2019advantage}, which also only evaluates the Q-function at state--action combinations that are present in the dataset:

\begin{equation} \label{eq:pi_loss}
\mathcal{L}_\pi(\phi) = \mathbb{E}_{\mathcal{D}} \left[ 
e^{\beta(Q_{\hat{\theta}} - V_\psi)} \log\pi_\phi(a|s) \right]
\end{equation}

where $\beta$ is an inverse temperature parameter which controls the trade-off between maximising the Q-function ($\beta\rightarrow\infty$) and staying close to the behaviour policy ($\beta\rightarrow0$) thereby supporting a spectrum of risk sensitivities.
AWR is based on a forward KL divergence regularisation pulling the policy towards actions observed in the data, pulling more strongly towards actions of higher estimated value.

\subsection{Lyapunov-inspired safety and PICNNs}

In classical control theory, one of the foundational tools for certifying stability and safety is the Lyapunov function. This is a scalar-valued function that decreases along system trajectories and provides a formal guarantee that the system will converge to a safe operating regime \cite{khalil1996nonlinear}. While constructing Lyapunov functions analytically for complex nonlinear systems is difficult, recent work has explored learning Lyapunov-like functions directly from data using structured NNs.

\begin{figure}[htb]
	\centering
	\includegraphics[width=0.49\textwidth]{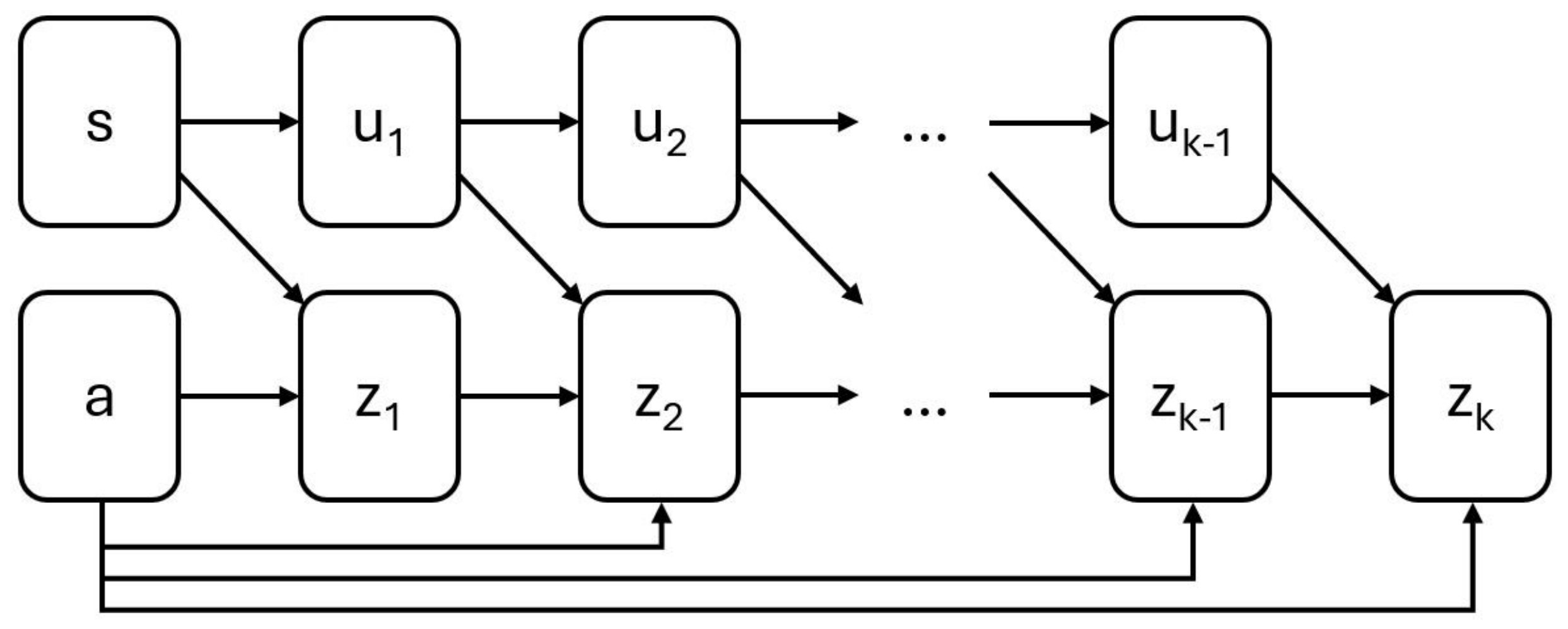}
	\caption{PICNN architecture adapted from \cite{amos2017input}. The network is convex in the action input $a$ while being expressive over the state $s$.}
	\label{fig:picnn_architecture}
\end{figure}

A promising approach in this direction is the use of PICNNs \cite{amos2017input}, which are feedforward architectures specifically designed to be convex in a subset of their inputs. Convexity plays a key role in optimisation and control, ensuring global optima and tractable inference. Standard NNs lack this property, limiting their applicability in structured prediction, inverse optimisation, and safe control tasks where convexity is desirable. PICNNs overcome this by constraining network weights and activations: all weights on the convex input path are non-negative, activation functions are convex and non-decreasing (e.g., ReLU, Softplus). To handle contextual variables such as system state, PICNNs also allow an additional non-convex input \( s \), while preserving convexity in the decision variable \( a \). The hidden layers of a PICNN are structured as:

\begin{equation}
    z_{i+1} = \phi(W_z^{(i)} z_i + W_a^{(i)} a + u_i(s))
\end{equation}

where \( a \) is the convex input and \( z_i \) are hidden activations, and \( u_i(s) \) is an unconstrained transformation of the state. This architecture is shown in Figure \ref{fig:picnn_architecture} 

In our work, we reinterpret PICNNs as deployment-time safety filters: by constructing a convex cost landscape over the action space, conditioned on the current state, we use the gradient of the PICNN to adjust potentially unsafe actions proposed by offline RL policies. This leverages the convexity of PICNNs for tractable correction, while preserving expressivity over the system state—a crucial requirement for safety-critical industrial control tasks. This approach can be considered as modelling a learned Lyapunov-like cost function upon which gradient descent is performed at test time to decrease the cost and guide the system towards safer operating regions over the course of the rollout.

\section{Benchmark environment} \label{sec:benchmark}
A good benchmark environment with corresponding datasets is required to explore RL for process control applications.
Whilst \cite{ahmed2025comparative} and \cite{bloor2024pc} recently proposed several benchmark problems for this domain, their case studies do not capture one of the most practically relevant and challenging scenarios: product grade transitions.
These transitions are particularly well-suited to RL, as they are often executed under close operator supervision and frequently involve manual interventions leading to rich historical datasets.
In this paper we introduce a new benchmark environment for product grade transitions, derived from a similar suspended polymerisation reactor example from literature \cite{maner1997polymerization}.

A simulation environment was developed based on a continuous stirred tank reactor (CSTR) undergoing an exothermic free-radical polymerisation reaction. The environment was implemented using the Gymnasium API to ensure compatibility with standard RL workflows. It incorporates detailed material and energy balances to realistically capture the nonlinear dynamics and control challenges inherent to industrial polymerisation processes.

\subsection{Reactor model}
\begin{figure}[htb]
    \centering
    \includegraphics[width=0.49\textwidth]{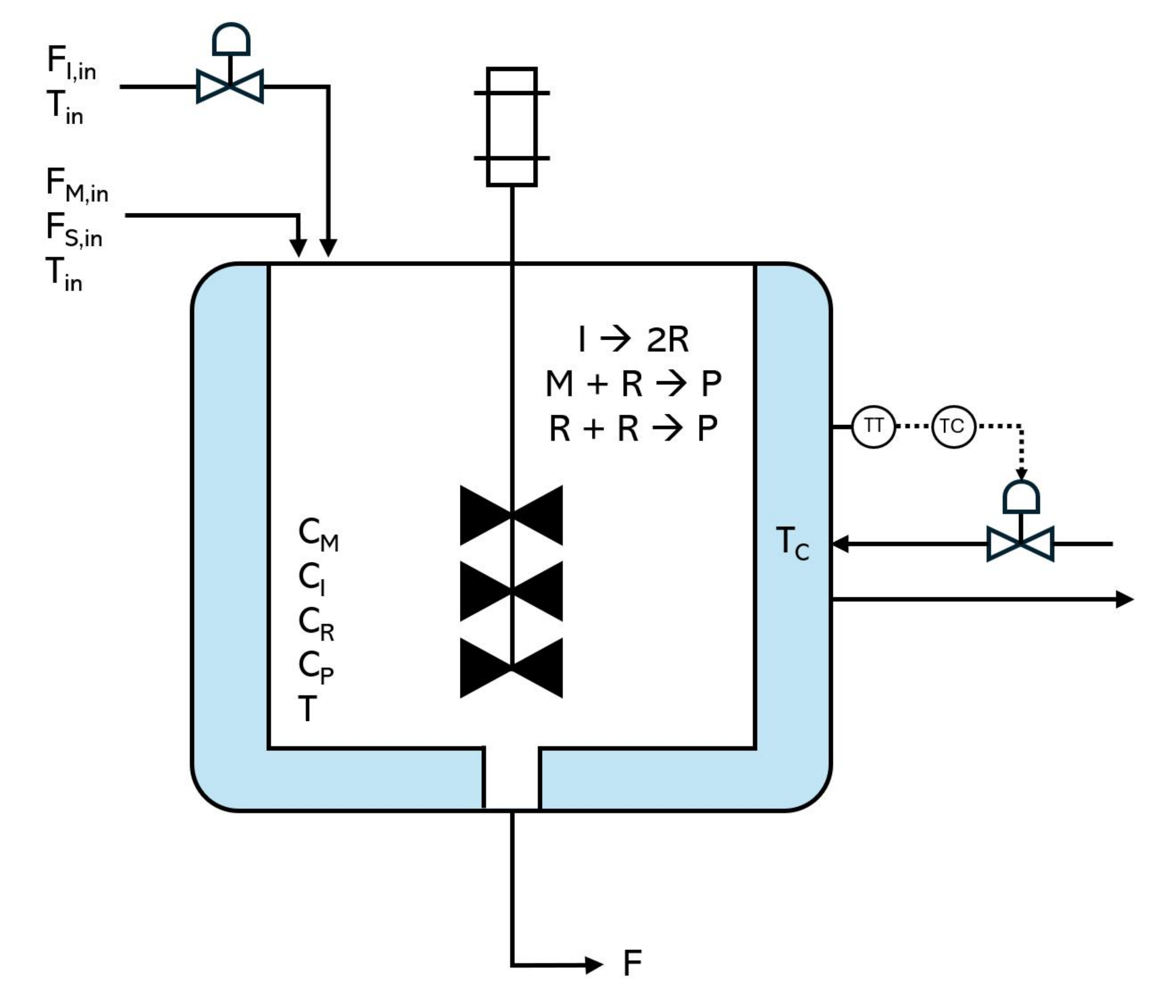}
    \caption{Schematic of the polymerisation CSTR environment. Control actions include the initiator feed rate ($F_{\text{M,in}}$) and cooling jacket temperature $T_C$. The reactor states include monomer concentration ($C_M$), initiator concentration ($C_I$), radical concentration ($C_R$), polymer concentration ($C_P$), and reactor temperature ($T$), where the latter two are also the controlled variables.}
    \label{fig:poly_cstr_schematic}
\end{figure}

A schematic of the CSTR is shown in Figure \ref{fig:poly_cstr_schematic}.
The CSTR is fed with a solvent (S) containing monomer species (M) and an initiator species (I) at temperature $T_\text{in}$. In the CSTR, the monomer and initiator react to form a polymer product (P) via a suspended lumped radical (R) mechanism. The CSTR outlet flowrate is assumed to be equal to the total inlet flowrate, maintaining a constant liquid filled reactor volume (1 m$^3$). Perfect mixing is assumed along with constant fluid density (1000 kg/m$^3$). Any gas phase effects and changes in viscosity due to polymer build-up are neglected.

The reaction pathway includes thermal auto-decomposition of the initiator to generate radicals (\ref{reaction:init}), propagation of monomer with radicals to form polymer chains (\ref{reaction:prop}), and bimolecular termination of radicals (\ref{reaction:term}). In this mechanism, all radical species (including free radical and propagating chain radicals) are represented with a single concentration variable for the lumped radicals, $C_R$. This approximation simplifies the kinetic model by avoiding the need to track individual chain lengths. The reaction set (\ref{reaction:init}--\ref{reaction:term}) describes the chemical reactions.

\begin{align}
    \text{Initiation:} \quad & \ce{I ->[k_i] 2R{\cdot}} \label{reaction:init}\\
    \text{Propagation:} \quad & \ce{M + R{\cdot} ->[k_p] P{\cdot}} \label{reaction:prop}\\
    \text{Termination:} \quad & \ce{R{\cdot} + R{\cdot} ->[k_t] P} \label{reaction:term}
\end{align}

Reaction kinetics are modeled using temperature-dependent Arrhenius expressions with constant pre-exponential factors and activation energies given in Table \ref{tab:poly_cstr_params}. All reactions are assumed to be kinetically controlled with no mass transfer limitations.

\begin{equation}
    k = A \exp\left(-\frac{E}{RT}\right)
\end{equation}

The reaction rates are defined as follows:
\begin{align}
    \text{Initiation:} \quad & r_i = k_i C_I \\
    \text{Propagation:} \quad & r_p = k_p C_M C_R \\
    \text{Termination:} \quad & r_t = k_t C_R^2
\end{align}

\subsection{Dynamics and numerical integration}
The reactor dynamics are governed by a system of coupled ordinary differential equations (\ref{ode1}--\ref{ode5}) using the parameters in Table \ref{tab:poly_cstr_params}. Heat effects are explicitly modelled, with an exothermic heat of reaction and thermal exchange with a cooling jacket. These are solved at each simulation step using SciPy's solveivp function with a stiff solver (BDF). The integration time step is set to 30 minutes to reflect typical control intervals for large-scale batch or semi-batch reactors. The simulation is terminated after 100 hours of operation unless otherwise specified.

\begin{align}
    \frac{dC_M}{dt} &= \frac{F}{V}\left(C^\text{in}_M - C_M\right) - r_p \label{ode1}\\
    \frac{dC_I}{dt} &= \frac{F}{V}\left(C^\text{in}_I - C_I\right) - r_i \\
    \frac{dC_R}{dt} &= \frac{F}{V}\left(C^\text{in}_R - C_R\right) + 2r_i - 2r_t \\
    \frac{dC_P}{dt} &= \frac{F}{V}\left(C^\text{in}_P - C_P\right) + r_p \\
    \frac{dT}{dt} &= \frac{F}{V}(T_f - T) - \frac{U A (T - T_c)}{\rho C_p V} - \frac{r_p \Delta H_{\text{rxn}}}{\rho C_p} \label{ode5}
\end{align}

Table \ref{tab:poly_cstr_params} gives the parameters used to construct the polymerisation reactor environment. The reaction is inspired by the polymerisation of vinyl acetate in benzene as presented in \cite{maner1997polymerization}, with some simplifications and rounding of parameters.

\begin{table}[htb]
	\centering
	\caption{Summary of reaction/reactor parameters used in the PolyCSTR environment}
	\label{tab:poly_cstr_params}
	\begin{tabular}{llll}
		\hline
		\textbf{Parameter} & \textbf{Description} & \textbf{Value} & \textbf{Unit} \\
		\hline
		\multicolumn{4}{c}{\textbf{Physical Constants}} \\
		\hline
		$R$ & Gas constant & 8.314 & J/mol/K \\
		$\rho$ & Density & 1000 & kg/m$^3$ \\
		$C_p$ & Heat capacity & 2000 & J/kg/K \\
		$U$ & Heat transfer coefficient & 500 & J/s/K \\
		$\Delta H_{\text{rxn}}$ & Heat of reaction & -100 & kJ/mol \\
		\hline
		\multicolumn{4}{c}{\textbf{Reactor Configuration}} \\
		\hline
		$V$ & Reactor volume & 1.0 & m$^3$ \\
		$F_S$ & Feed solvent rate & 80 & kg/h \\
		$F_M$ & Feed monomer rate & 100 & kg/h \\
		$T_f$ & Feed temperature & 350 & K \\
		\hline
		\multicolumn{4}{c}{\textbf{Kinetics}} \\
		\hline
		$A_{\text{init}}$ & Pre-exponential factor (initiation) & $1 \times 10^9$ & 1/s \\
		$E_{\text{init}}$ & Activation energy (initiation) & 125 & kJ/mol \\
		$A_{\text{prop}}$ & Pre-exponential factor (propagation) & $4 \times 10^4$ & m$^3$/mol/s \\
		$E_{\text{prop}}$ & Activation energy (propagation) & 25 & kJ/mol \\
		$A_{\text{term}}$ & Pre-exponential factor (termination) & $1 \times 10^6$ & m$^3$/mol/s \\
		$E_{\text{term}}$ & Activation energy (termination) & 15 & kJ/mol \\
		\hline
		\multicolumn{4}{c}{\textbf{Control Inputs}} \\
		\hline
		$F_I$ & Initiator feed rate & 0.0 - 2.5 & kg/h \\
		$T_c$ & Coolant temperature & 300 - 355 & K \\
		\hline
		\multicolumn{4}{c}{\textbf{Simulation Settings}} \\
		\hline
		$\Delta t$ & Time step & 0.5 & h \\
		$T_{max}$ & Max simulation time & 100 & h \\
		\hline
	\end{tabular}
\end{table}

\subsection{RL environment implementation}
The reactor model is embedded in a Gymnasium environment and key environment parameters are listed in table \ref{tab:environment}.

\begin{table}[htb]
	\centering
	\caption{Polymerisation reactor Gymnasium environment characteristics.}
	\label{tab:environment}
	\begin{tabular}{llll}
		\cline{2-4}
		\addlinespace[0.25em]
		& \textbf{Variable} & \textbf{Description} & \textbf{Unit} \\
		\addlinespace[0.25em]
		\cline{2-4}
		\addlinespace[0.25em]
		\multirow{9}{*}{\textbf{States}} 
		& $C_M$ & Monomer concentration & kg/m$^3$ \\
		& $C_P$ & Polymer concentration & kg/m$^3$ \\
		& $T$ & Reactor temperature & K \\
		& $e_{C_P}$ & Polymer SP tracking error & kg/m$^3$ \\
		& $e_T$ & Temperature SP tracking error & K \\
		& $\Delta e_{C_P}$ & Change in polymer SP tracking error & kg/m$^3$ \\
		& $\Delta e_T$ & Change in temperature SP tracking error & K \\
		& $F_{\text{init}}$ & Current initiator feed rate & kg/h \\
		& $T_c$ & Current coolant temperature & K \\
		\addlinespace[0.25em]
		\cline{2-4}
		\addlinespace[0.25em]
		\multirow{2}{*}{\textbf{Actions}} 
		& $\Delta F_{\text{init}}$ & Change in initiator feed rate & kg/h \\
		& $\Delta T_c$ & Change in coolant temperature & K \\
		\addlinespace[0.25em]
		\cline{2-4}
		\addlinespace[0.25em]
		\textbf{Reward} & $r(s,a)$ & Setpoint tracking with stability bonus (Eq.~\ref{eq:reward}) & -- \\
		\addlinespace[0.25em]
		\cline{2-4}
	\end{tabular}
\end{table} 

RL agents interact with the environment via two continuous control actions: the initiator feed rate (kg/h) and the cooling jacket temperature (K). These are bounded within operationally realistic ranges: 0.0--2.5 kg/h for the initiator and 300--350 K for the coolant temperature. The observation space includes three measured process variables: monomer concentration, polymer concentration, and temperature of the outlet. Note that this leaves important simulation state variables such as initiator and radicals concentration invisible to the agent. Noise is added on the process variables measurements to add complexity and realism. Both temperature and polymerisation have setpoints, and the observation space is augmented with error (2x) and error rate of change (2x) variables making for a total of 7 process related observations. To facilitate learning directionally correct strategies, a delta-action approach is taken. Thus, the state space is further augmented with the current absolute values of the action variables, bringing the total observation space to 9 continuous variables. The action space is given by the change in initiator feed and change in coolant temperature with respect to the current situation.

\subsubsection{Rewards}

To guide the agent toward effective and safe grade transitions, we design a reward function that encourages progress toward the desired setpoint while penalising unsafe behaviour.
The primary reward signal is based on error reduction: the agent is rewarded for decreasing the absolute deviation in concentration and temperature from their respective targets between time steps.
This difference-in-error formulation incentivises consistent improvement rather than absolute accuracy at any single step.
To further stabilise control near the target, we introduce a setpoint proximity bonus: if both concentration and temperature remain within predefined thresholds (2.0 and 1.0 units, respectively), the agent receives an additional reward scaled linearly by how close the state is to the setpoint.
This encourages the agent not only to reach the desired operating region but to remain there with minimal deviation.
Finally, to ensure safety, a hard penalty of $-1000$ is applied if the reactor temperature drops below $-50^\circ C$, representing a thermal runaway condition.
This structure balances short-term progress, long-term stability, and safety-critical constraints—key priorities in industrial process control.
The reward function is hereby defined as:

\begin{equation} \label{eq:reward}
r_t =
\begin{cases}
    -1000, & \text{if } e_T < -50 \quad \text{(thermal runaway)} \\[1ex]
    \Delta e_t + b_t, & \text{otherwise}
\end{cases}
\end{equation}

\noindent\text{where:}
\begin{align}
\Delta e_t &= \left( |e_{C_P}^{t-1}| + |e_T^{t-1}| \right) - \left( |e_{C_P}^t| + |e_T^t| \right) \quad \text{(error reduction)} \\[1ex]
b_t &= 
\begin{cases}
10 \cdot \left( 1 - \max\left( \frac{|e_{C_P}^t|}{\tau_{C_P}}, \frac{|e_T^t|}{\tau_T} \right) \right), & \text{if } |e_{C_P}^t| < \tau_{C_P} \text{ and } |e_T^t| < \tau_T \\[1ex]
0, & \text{otherwise}
\end{cases}
\end{align}

\noindent\text{with thresholds:} \quad \( \tau_{C_P} = 2.0 \) kg/m\(^3\), \quad \( \tau_T = 1.0 \) K

\subsubsection{Costs}

In addition to the reward signal used for policy optimisation, we define a separate cost function based on the sum of absolute setpoint tracking errors:

\begin{equation}
c_t = |e_{C_P}^t| + |e_T^t|
\end{equation}

This cost serves a distinct role from the reward. While the reward encourages immediate progress and incentivises reaching the setpoint region, the cost is designed to be always non-negative, enabling its use as a proxy for system stability—analogous to a Lyapunov function. In our architecture, this cost function is leveraged by the safety filter to perform deployment-time correction: the PICNN models a state-conditioned surrogate cost landscape, and its gradient is used to adjust unsafe actions proposed by the offline RL policy in a direction that locally reduces this cost. Defining the cost separately allows the reward to remain flexible (e.g., for shaping learning signals), while the cost is tightly aligned with operational safety and long-term control objectives. More generally, in industrial process control settings, such costs could include not only setpoint deviations but also absolute constraint violations, such as pressure limits, flow bounds, or temperature thresholds.

\subsection{Control scenarios}
Three scenarios are considered for the polymerisation reactor control.
\begin{enumerate}
	\item \textbf{Startup}. In startup mode the reactor is initialised at equilibrium from a temperature and composition view, but without any initiator flow. Without initiator, there are no radicals and there can be no polymerisation. The objective in this scenario is to go from a polymer concentration of $0$ kg/m$^3$ to $100$ kg/m$^3$, while maintaining reactor temperature at $350$ K.
	\item \textbf{Grade change down}. In this scenario the reactor the starting point is equilibrium operation at $350$ K and $100$ kg/m$^3$ of polymer. The objective is to change the operation to a new setpoint of $355$ K and $90$ kg/m$^3$ of polymer. At the higher temperature with lower conversion the chain length distribution will skew towards the shorter chains yielding a lower density polymer.
	\item \textbf{Grade change up}: This is the reverse of scenario 2, moving back towards $350$ K and $100$ kg/m$^3$ polymer.
\end{enumerate}

To generate offline data for training and evaluation, PI controllers were implemented to regulate polymer concentration and reactor temperature to their predefined setpoints through manipulation of the initiator feed flow and cooling jacket temperature respectively. The initialisation is such that there is no p-action kick from the setpoint change. The PI algorithm gain parameters (K$_p$ and K$_i$) are uniformly sampled from a factor 10 range, giving variation between rollouts. Anti-windup logic is applied to prevent further integral term accumulation when actuators are saturated.

Each scenario is rolled out 100 times in simulation, for a duration of 100 hours with a 30 min sample frequency. This creates a dataset of 20,000 samples per scenario across 100 different PI tuning parameter combinations. Note that the reaction kinetics and temperature controller strength are such that there is a real risk of a runaway reaction if the temperature get too high in the presence of sufficient initiator, creating an interesting control challenge. With more conservative controller settings, the reactor won't reach the setpoint within the 100 hours of simulation time. With more aggressive settings the reaction may runaway uncontrollably in the more challenging startup and grade change down scenarios. By comparison, the grade change up is comparatively simpler as the temperature is reduced rather than increased. Thermal runaway would occur at a reactor temperature of approximately $365$ K, depending on the exact concentration of monomer and initiator.

% See the picture below for the polymer concentration across the rollouts for each scenario.
The challenge for the RL agent is to distill a policy combining the best parts of all the trajectories supported by the data.
Aggressive where it can be, careful where it needs to be.
The non-linear nature of reinforcement learning agents allows for variation of the strategy with respect to reactor conditions that cannot be captured in a fixed PI controller strategy.

\section{Offline learning} \label{sec:offline}

\subsection{Behaviour cloning}

We trained a BC agent as a supervised learning baseline using a dataset of state-action pairs collected from the PolyCSTR environment. The policy is represented by a multilayer perceptron (MLP) with two hidden layers, each containing 256 units and ReLU activations. The network maps observed states to actions under a Gaussian policy, where the mean and standard deviation are predicted by the network. The policy was optimised using the Adam optimiser with a learning rate of $0.0003$. Training was conducted for 200 epochs with a batch size of 512, using the full offline dataset as a replay buffer. The output actions were normalized and scaled to lie within the bounds of $[-1, +1]$, with a maximum action magnitude set to 1.0. This BC agent serves as a strong supervised learning baseline, capturing expert-like behavior directly from demonstration data without any explicit RL objective.

\subsection{IQL agent training}

We implemented IQL agents using NN architectures for the Q-function, value function, and policy networks.
Each of these networks was parameterised as a fully connected multilayer perceptron (MLP) detailed below.

\begin{itemize}
    \item Q-network: The Q-function was modeled using a feedforward MLP with 2 hidden layers, each containing 256 units and using ReLU activation functions. The output layer produced a single scalar Q-value estimate for each state--action pair.
    \item Value network: The state-value function was modeled similarly with 2 hidden layers and 256 units per layer, also using ReLU activations. The output was a scalar estimate of the expected return from a given state.
    \item Policy network: The policy was represented as a stochastic policy with a Gaussian head, using an MLP with 2 hidden layers, 256 units each, and ReLU activations. The output layer produced the mean and log standard deviation of the action distribution.
\end{itemize}

The IQL networks were trained on the offline polymerisation CSTR datasets.
Training was conducted using mini-batches of size 512.
The temporal discount factor was set to 0.9.
Target networks were updated using a soft update rule with a coefficient of 0.05.
The expectile parameter $\tau$ used in the expectile regression loss for the value function was set to 0.9.
The temperature $\beta$ governing the weight of Q-values in policy extraction was set to 5.
Training was performed over 2000 epochs using a replay buffer containing 20100 transitions.
Learning rate decay was implemented using a cosine annealing schedule.

\subsection{Learning PICNN cost models}

A key challenge in deploying offline RL policies in real-world environments lies in ensuring practical safety whilst ensuring policy competence.
One promising strategy involves learning a cost function that captures constraint violations, and using its gradient to refine the policy's action outputs at test time.
However, standard NN cost models are typically non-convex in the action space, leading to unreliable gradients during correction which can be particulalry problematic in safety-critical domains.

In this work, we propose to address this issue by leveraging PICNNs as cost models.
The PICNNs were trained to ensure convexity in the action space, whilst ensuring full expressivity over the state space.
This structural property ensures that the learned cost function $c(s, a)$ is convex in $a$ for every state $s$, enabling stable and predictable gradient-based corrections.
To the best of our knowledge, this is the first application of PICNNs as cost models for real-time action refinement in safe RL deployment.
we demonstrate that enforcing action-convexity in the cost model leads to smoother cost landscapes, more reliable descent directions, and ulitimately, safer policy behaviour during deployment.

PICNN cost models were trained and compared to standard NNs. For comparison, the standard NN and the PICNN cost models were trained using the same paramters and architecture, with the only difference being that the PICNN cost model was trained to ensure convexity in the action space.

\begin{table}[htb]
	\centering
	\caption{PICNN and standard NN cost model training parameters.}
	\label{tab:picnn_training}
	\begin{tabular}{l l}
		Parameter & Value \\
		\hline
		No. hidden layers & 2 \\
		No. hidden nodes & 64 \\
		Activation function & Softplus \\
		Max no. epochs & 1000 \\
		Minibatch size & 512 \\
		Learning rate & 0.001 \\
		Optimiser & Adam \\
		Loss function & MSE \\
	\end{tabular}
\end{table}

\section{Online action correction} \label{sec:online}

We propose a deployment-time safety mechanism that refines actions from an offline policy using a learned cost model. Specifically, we model a state-conditioned cost function \( \hat{c}_\phi(s, a) \) using a PICNN, which is guaranteed to be convex in the action \( a \) and expressive in the state \( s \). Given that convexity in \( a \) ensures a unique global minimum, we can apply principled optimisation methods to correct unsafe or suboptimal actions proposed by the policy.

Let \( a_0 = \pi(s) \) be the original action suggested by the trained policy. The goal is to compute a corrected action \( a^* \) that reduces the cost \( \hat{c}_\phi(s, a) \) while remaining computationally feasible for real-time control.

\paragraph{Gradient Descent Update.} The simplest update rule uses the gradient of the cost with respect to the action:
\begin{equation}
    a_{k+1} = a_k - \eta \nabla_a \hat{c}_\phi(s, a_k)
\end{equation}
where \( \eta > 0 \) is a fixed or adaptive step size. This update moves in the direction of steepest descent and is guaranteed to reduce the cost under convexity assumptions.

\paragraph{Newton Step Update.} When the second derivative (Hessian) is available and the cost function is twice-differentiable in \( a \), a more efficient update is given by:
\begin{equation}
    a_{k+1} = a_k - \left( \nabla^2_a \hat{c}_\phi(s, a_k) \right)^{-1} \nabla_a \hat{c}_\phi(s, a_k)
\end{equation}
This Newton step can offer faster convergence, especially near the minimum, by adjusting the step direction and size using local curvature.

While standard NNs often yield highly non-convex loss landscapes, PICNNs enforce convexity in action space through architectural constraints (non-negative weights, convex activations), ensuring that updates reliably move toward a well-defined minimum. This makes the correction process stable, interpretable, and efficient which are key properties for deployment in industrial process control systems.

At each timestep during online deployment, the system observes the current state, queries the policy for an action, and refines this action using the gradient (or Newton) steps over the PICNN-modeled cost surface. This safety layer operates without modifying the policy network and without requiring exploration, making it compatible with any offline-trained policy and suitable for constrained real-time settings.

\begin{algorithm}[htb]
\caption{Online Deployment with Action Correction via PICNN}
\label{alg:picnn}
\begin{algorithmic}[1]
\State \textbf{Inputs:} Trained policy \( \pi \), trained cost model \( \hat{c}_\phi(s, a) \), environment \( \mathcal{E} \), step size \( \eta \)
\State Initialise state \( s_0 \)
\For{each timestep \( t = 0, 1, 2, \dots \)}
    \State Observe current state \( s_t \) from environment
    \State Compute proposed action \( a_t \gets \pi(s_t) \)
    \State Compute gradient \( \nabla_a \hat{c}_\phi(s_t, a_t) \)
    \If{Newton step enabled}
        \State Compute Hessian \( H_t \gets \nabla^2_a \hat{c}_\phi(s_t, a_t) \)
        \State Correct action: \( a_t \gets a_t - H_t^{-1} \nabla_a \hat{c}_\phi(s_t, a_t) \)
    \Else
        \State Correct action: \( a_t \gets a_t - \eta \cdot \nabla_a \hat{c}_\phi(s_t, a_t) \)
    \EndIf
    \State Optionally clip or project \( a_t \) to valid action bounds
    \State Execute action \( a_t \) in environment: \( s_{t+1}, r_t \gets \mathcal{E}(s_t, a_t) \)
\EndFor
\end{algorithmic}
\end{algorithm}

\section{Results} \label{sec:results}

\subsection{PICNN cost model analysis}

\begin{figure}[htb]
	\centering
	\includegraphics[width=0.49\textwidth]{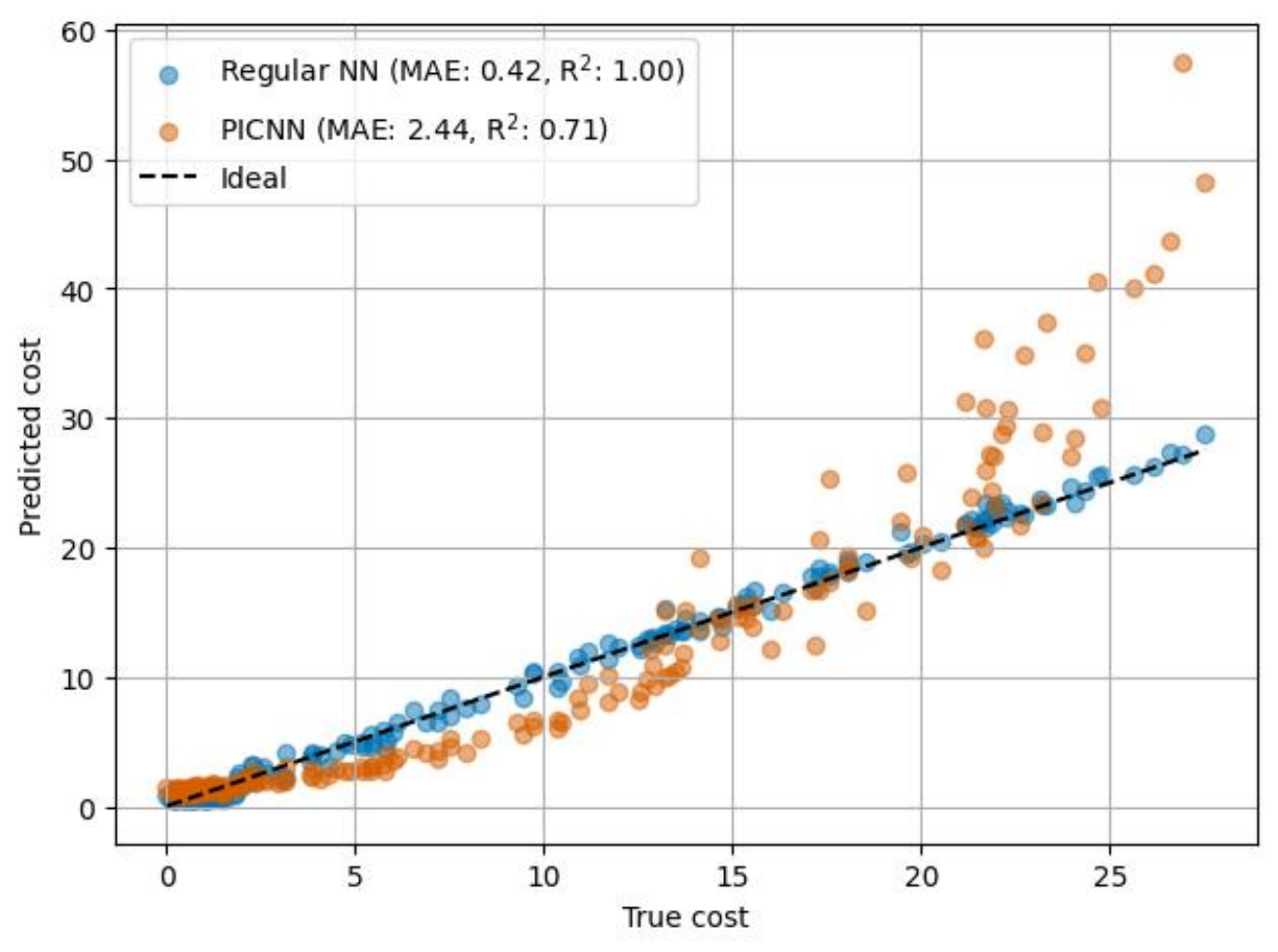}
	\caption{Comparison of standard NN and PICNN cost model predictions on training data from the polymerisation CSTR grade change up scenario. The PICNN trades predictive performance for convexity.}
	\label{fig:nn_grade_up}
\end{figure}

\begin{figure}[htb]
	\centering
	\includegraphics[width=0.49\textwidth]{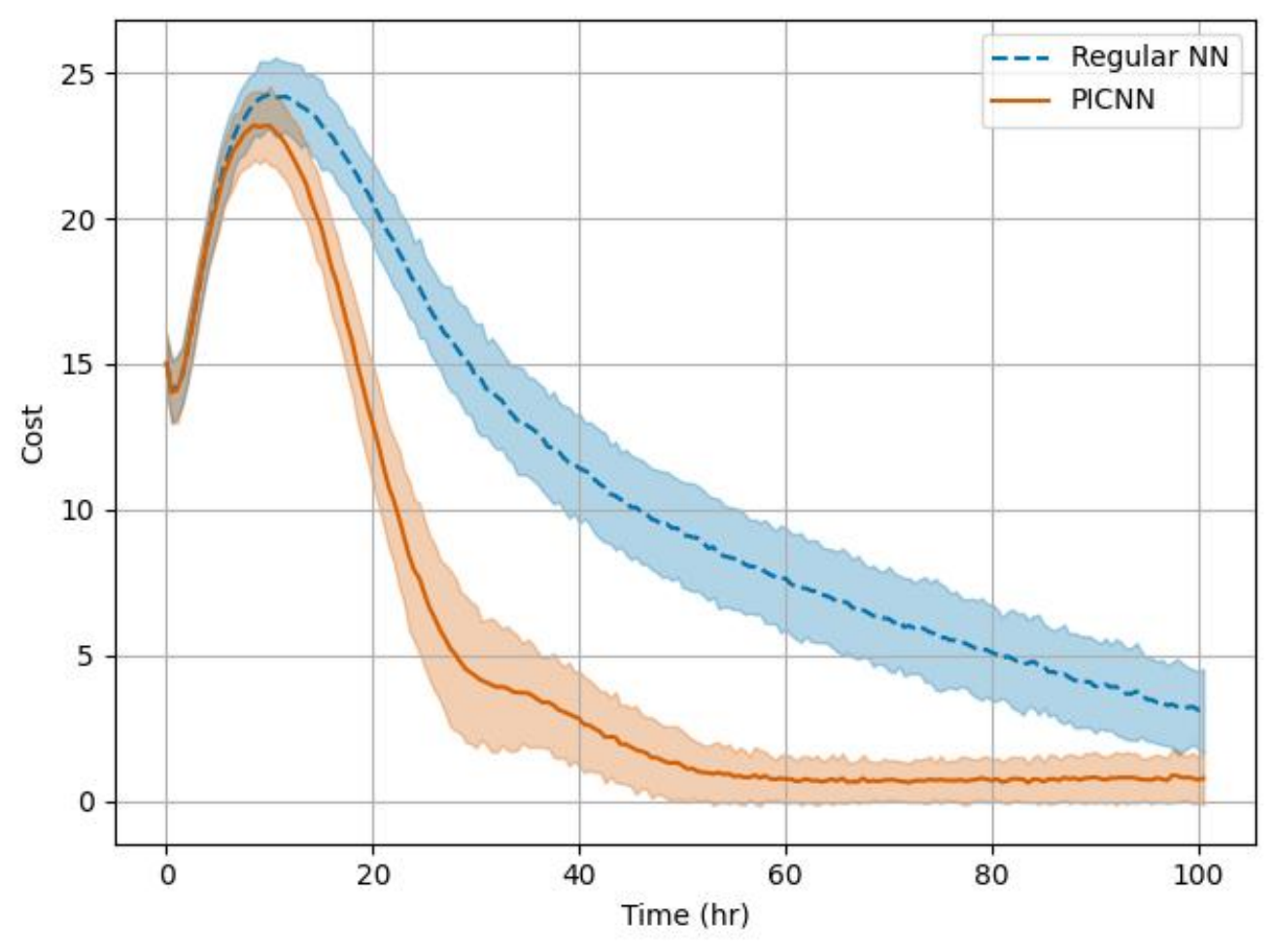}
	\caption{Comparison of environment cost feedback during rollouts of the polymerisation CSTR grade change up scenario controlled an IQL agent with online gradient step corrections on two cost models: a regular NN, and a PICNN. The line shows the mean cost, whilst the shaded area shows 2 standard deviations over 100 rollouts.}
	\label{fig:cost_rollouts}
\end{figure}

Figure \ref{fig:nn_grade_up} presents a comparison of predicted versus true costs for the grade change up scenario, evaluated on a test set.
The regular NN exhibits strong predictive accuracy, with a mean absolute error (MAE) of 0.47 and a coefficient of determination (R$^2$) of 0.99, closely tracking the ideal (i.e., predicted cost equals true cost) line.
In contrast, the PICNN demonstrates a higher MAE of 2.20 and a notably lower R$^2$ of 0.73, indicating reduced accuracy in pointwise cost estimation.
This performance gap is particularly pronounced for samples with higher true cost values, where the PICNN tends to overpredict.
The discrepancy reflects the trade-off between flexibility and structure: while the PICNN imposes convexity with respect to actions, it also constrains model expressiveness, potentially limiting its fit to complex cost surfaces.
Despite this lower regression accuracy, as shown in subsequent rollout experiments (Figure \ref{fig:cost_rollouts}), the PICNN's structural bias can still translate to more reliable online control through stable gradient corrections.
This highlights the tradeoff between enforcing convexity constraints and achieving optimal predictive accuracy.

Figure \ref{fig:cost_rollouts} shows the rollout cost comparison over time for the grade change up scenario, where an IQL agent performs online control using gradient-based corrections informed by a learned cost model. The comparison is between the regular NN (dashed blue line) and the PICNN (solid orange line). Shaded regions indicate two standard deviations from the mean cost over 100 rollouts. While both models initially yield similar cost trajectories, the PICNN leads to more consistent and stable cost reduction over time, particularly at lower costs. This improved convergence behavior highlights the benefit of action convexity in the cost model, enabling more reliable gradient-based corrections during online deployment.

These results highlight a key insight: accurate cost prediction does not guarantee effective control. The PICNN, though less precise in estimating absolute cost values (Figure \ref{fig:nn_grade_up}), provides more consistent and reliable gradients for online optimisation (Figure \ref{fig:cost_rollouts}), likely due to its convexity constraint in the action space. This structural bias appears to regularise the gradient landscape, enabling the agent to more effectively descend toward low-cost trajectories, especially over longer horizons.

\begin{figure}[htb]
	\centering
	\includegraphics[width=0.7\textwidth]{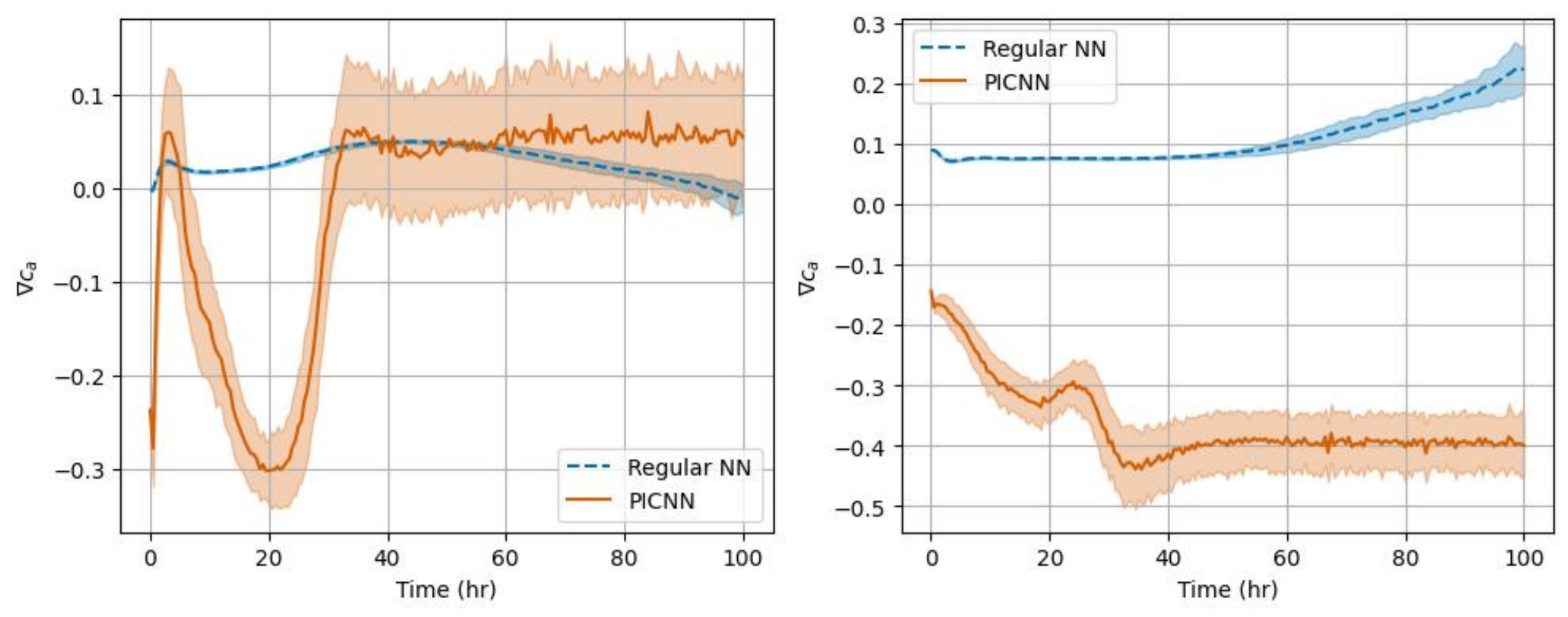}
	\caption{Comparison of gradients during rollouts of the polymerisation CSTR grade change up scenario controlled by a BC agent with online gradient step corrections on two cost models: a regular NN, and a PICNN. Left: cost gradient with respect to initiator feed action. Right: cost gradient with respect to coolant temperature action. The line shows the mean gradient magnitude, whilst the shaded area shows 2 standard deviations over 100 rollouts.}
	\label{fig:gradients}
\end{figure}

To further investigate the observed difference in rollout performance between the PICNN and NN cost models, Figure \ref{fig:gradients} compares the gradients of the predicted cost with respect to each action dimension (initiator feed, left, and coolant temperature, right) over the course of the rollouts. The regular NN produces gradients that are initially small and relatively flat, but increase in magnitude and variability over time, especially for the coolant temperature. This growing instability may lead to overly aggressive or misdirected corrections later in the rollout. In contrast, the PICNN exhibits a more structured and bounded gradient profile. For the initiator feed, the PICNN responds quickly with strong early gradients, followed by a steady stabilisation. For the coolant temperature, the PICNN gradients remain consistently negative and smoother throughout, providing more reliable directional information for control.

Interestingly, the gradients predicted by the two models for coolant temperature not only differ in magnitude but often point in opposite directions: PICNN consistently suggests decreasing the coolant temperature, while the Regular NN trends upward. This directional disagreement implies a fundamental mismatch in the optimisation signals, which can lead to divergent behaviours during control. Moreover, the NN gradients are generally smaller in magnitude across both action dimensions, indicating a flatter cost surface with respect to the actions. This may limit the effectiveness of gradient-based action refinement, as updates derived from such weak gradients result in negligible policy improvement. Despite the NN achieving a better overall cost prediction fit (Figure \ref{fig:nn_grade_up}), this flatness suggests that the model's cost surface may be overly smooth or under-sensitive to action changes. This, in turn, hints at the NN relying more heavily on its modelling capacity with respect to state features rather than action sensitivity.

By contrast, the PICNN's convexity constraint limits its expressiveness in fitting arbitrary cost functions, but encourages stronger, more informative gradients with respect to actions. This structural bias may shift some modelling flexibility into the state representation, effectively balancing action sensitivity and state expressiveness. The result is a more useful gradient field for online control even if it comes at the cost of slightly reduced prediction accuracy. These differences support the hypothesis that structural inductive biases like convexity can improve the utility of learned models in control settings by promoting more stable, interpretable, and effective gradients over the course of long-horizon decision making.

\subsection{Control performance}

\subsubsection{Dataset generation}

\begin{figure}[htb]
	\centering
	\includegraphics[width=0.7\textwidth]{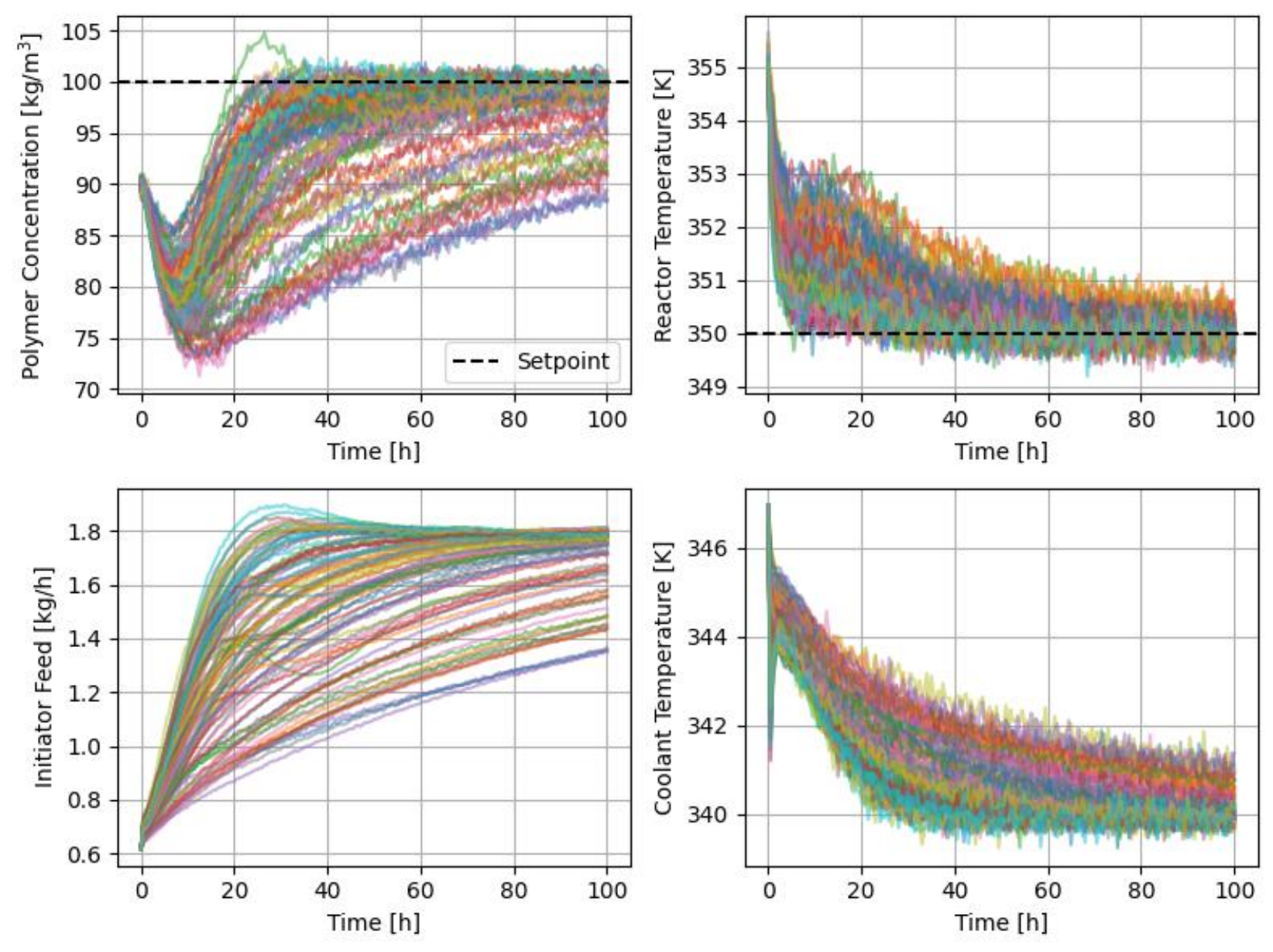}
	\caption{100 rollouts comprising the polymerisation CSTR grade up dataset.}
	\label{fig:data}
\end{figure}

Figure \ref{fig:data} illustrates the closed-loop behaviour of the polymerisation CSTR during the upward grade transition scenario. This scenario proved to be the most difficult of the 3 CSTR scenarios and so the results are presented here whilst the other scenario results are demonstrated in Sections \ref{sec:su_results} and \ref{sec:gd_results}. The dataset was generated using a population of PI controllers, each with randomly sampled tuning parameters to encourage diversity in trajectory behaviors. This approach simulates the type of variability and suboptimality that is common in industrial process operations, providing a realistic and heterogeneous dataset suitable for offline reinforcement learning.

The top-left panel shows the evolution of polymer concentration over time, with the desired setpoint indicated by a dashed black line. The system exhibits significant initial undershoot followed by a slow recovery, with varying degrees of convergence quality across trajectories. The spread in final concentration values highlights the impact of differing PI gains, with some controllers achieving near-setpoint regulation while others remain below the target.

The top-right panel depicts the reactor temperature response. All trajectories exhibit an initial overshoot and gradual settling near the target temperature, reflecting the thermal inertia of the system and the influence of controller tuning on transient dynamics.

In the bottom-left panel, the initiator feed rate shows rapid ramp-up behaviour with clear differences in aggressiveness and steady-state values, again due to tuning variation. This reflects how control input diversity is implicitly encoded into the training data through controller sampling.

Lastly, the bottom-right panel shows the coolant temperature, which similarly exhibits settling dynamics and trajectory variance. The consistency of temperature regulation across trajectories despite variation in control actions suggests that the underlying dynamics are well-behaved but sensitive to control design.

Overall, this dataset captures a broad spectrum of process responses and control actions, making it well-suited for training data-driven cost models and evaluating policy improvement techniques in offline reinforcement learning settings.

\subsubsection{Behaviour cloning}

\begin{figure}[htb]
	\centering
	\includegraphics[width=0.7\textwidth]{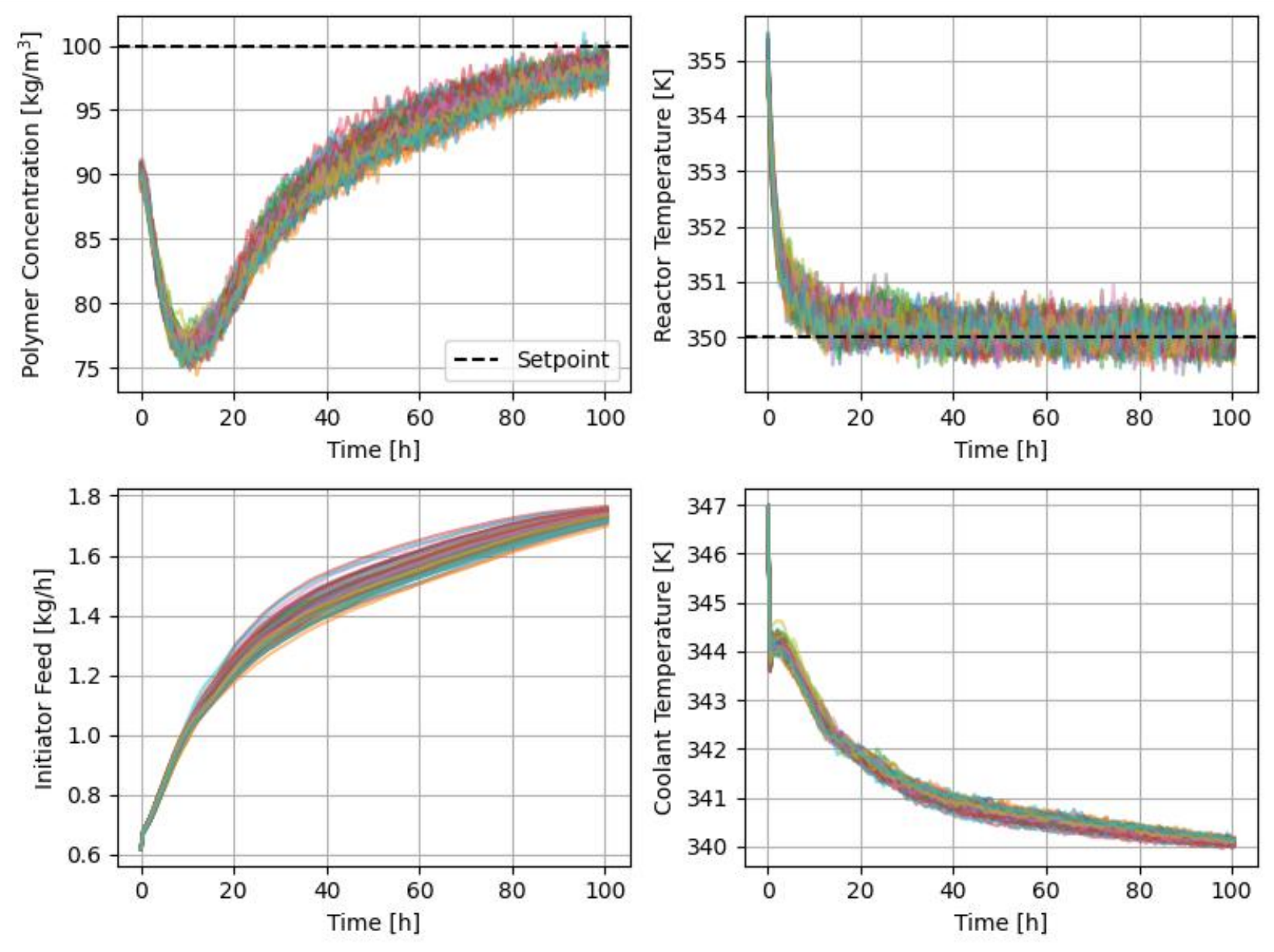}
	\caption{BC agent rollouts on the grade up scenario.}
	\label{fig:bc}
\end{figure}

Figure \ref{fig:bc} shows the closed-loop system behaviour under control by the BC policy trained on the diverse PI-generated dataset described earlier. While the agent is able to reproduce average-case control behaviour reasonably well, several key shortcomings are apparent, particularly in the polymer concentration dynamics.

Compared to the original PI controlled dataset, the BC policy exhibits a consistent tendency toward sluggish correction of polymer concentration, as shown in the top-left panel. While most trajectories eventually approach the setpoint, the rate of convergence is noticeably slower and the transient undershoot more prolonged. This reflects the policy's bias toward conservative control actions, likely stemming from the dominance of suboptimal (slow-reacting) behaviour in the training data. This skew is consequential, as it negatively impacts the policy's cumulative reward and explains its low performance scores in Table \ref{tab:results}.

The reactor temperature (top-right), initiator feed rate (bottom-left), and coolant temperature (bottom-right) dynamics are more consistent and display improved regularity compared to the training data. In particular, the control inputs show reduced variability, indicating that the BC agent has learned a smoothed policy that largely avoids erratic behaviour. However, this smoothing effect comes at the cost of reactivity, especially in the polymer concentration loop where timely intervention is critical for grade change performance.

Overall, while the BC policy captures the average tendencies of the training data, it lacks the adaptability required for faster or more precise responses. This highlights a key limitation of purely supervised offline policies in process control settings where asymmetric control quality in the dataset can propagate directly into suboptimal closed-loop behavior.

\subsubsection{BC with online gradient-based action corrections}

\begin{figure}[htb]
	\centering
	\includegraphics[width=0.7\textwidth]{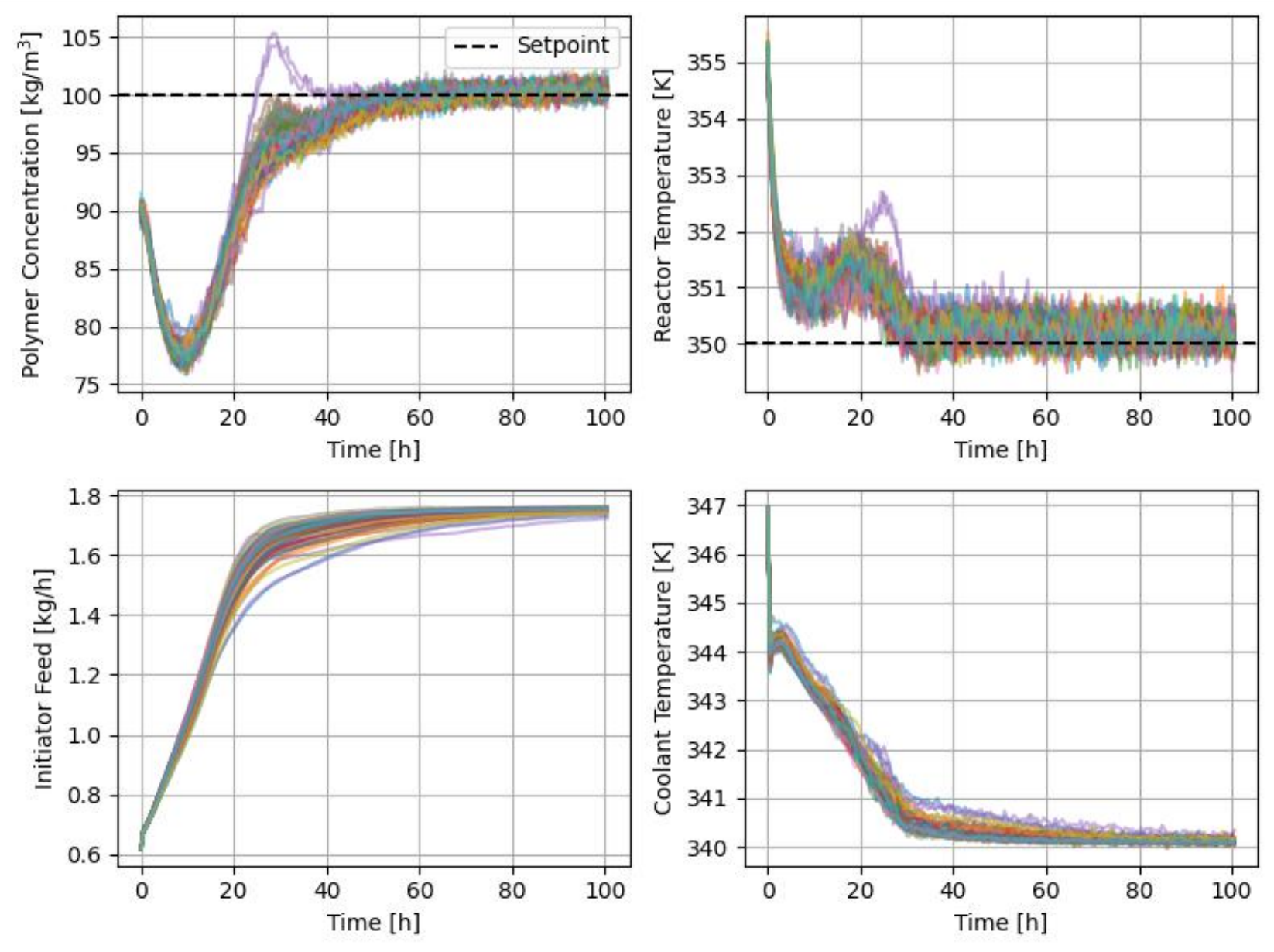}
	\caption{BC agent with online gradient-based action corrections on the grade up scenario.}
	\label{fig:bcp}
\end{figure}

Figure \ref{fig:bcp} presents the closed-loop performance of the same BC agent augmented with online gradient-based action corrections using a PICNN cost model. This hybrid approach leverages the BC policy for baseline control actions, which are then refined in real time via gradient steps that minimise the learned cost, effectively steering the system toward safer process control.

The most notable improvement is observed in the polymer concentration (top-left). Compared to Figure \ref{fig:bc}, the sluggishness seen under pure BC control has been largely eliminated. Trajectories now exhibit faster convergence to the setpoint, with significantly reduced steady-state error. In many cases, the system reaches the target concentration within the first 40 hours showing a substantial improvement in responsiveness. This comes at the cost of some overshoot in early transients, a side effect of the more aggressive corrections applied via the gradient-based updates. It was important to tune the step size of the gradient descent to balance responsiveness and stability, as too large a step could lead to larger overshoots and instability.

A similar pattern is visible in the reactor temperature (top-right), which now exhibits sharper initial transients but improved settling behavior overall. Importantly, the system remains within safe operational bounds across all trajectories, demonstrating the stability and safety-preserving nature of the cost model's convex structure.

The control inputs, initiator feed (bottom-left) and coolant temperature (bottom-right), reflect this increased actuation agility. The initiator feed trajectories ramp up more rapidly and saturate closer to the nominal upper limit, suggesting that the cost model guides the corrections toward energetically efficient, yet assertive control adjustments. Coolant temperature responses show slightly more variability than under pure BC control but remain stable and well-regulated.

Overall, the addition of PICNN-based gradient corrections transforms the BC agent's behavior from passively average to actively competent with a 14$\times$ score increase (Table \ref{tab:results}). While overshoots are introduced in the pursuit of faster setpoint tracking, the tradeoff appears favourable, especially in light of the consistent safety and marked performance gains. This result underscores the utility of convex cost models not only for offline policy evaluation, but also for safe and targeted online policy refinement.

\subsubsection{IQL agent performance}

\begin{figure}[htb]
	\centering
	\includegraphics[width=0.7\textwidth]{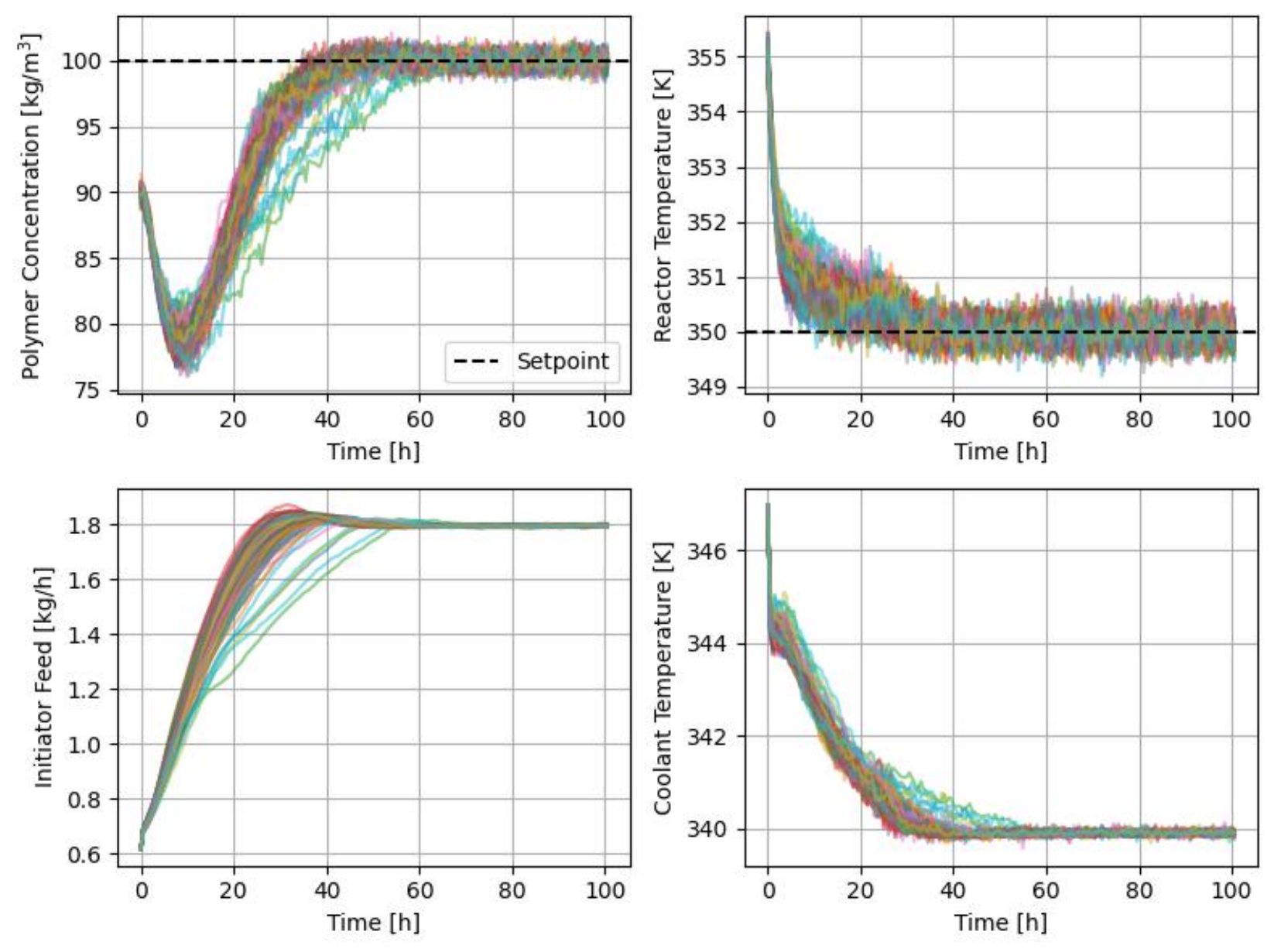}
	\caption{IQL agent rollouts on the grade up scenario.}
	\label{fig:iql}
\end{figure}

Figure \ref{fig:iql} presents the closed-loop performance of an IQL policy deployed in the same grade-up scenario. Among all controllers evaluated so far, this pure offline RL approach yields the most consistent and performant behavior, setting a clear benchmark for comparison.

The polymer concentration trajectories (top-left) demonstrate rapid and reliable convergence to the setpoint with minimal overshoot and negligible steady-state error. Unlike the BC and BC+ variants, IQL achieves both responsiveness and precision without compromising stability. The variability between trajectories is markedly reduced, indicating strong policy generalisation across initial conditions and process disturbances.

Reactor temperature (top-right) regulation is equally impressive. All trajectories settle close to the setpoint with minimal oscillation or noise amplification, suggesting that the learned value function effectively encoded not only the target behavior but also the cost of unnecessary actuation. The control effort is smooth and well-coordinated.

The initiator feed (bottom-left) and coolant temperature (bottom-right) profiles further underscore the quality of the policy. Feed trajectories rise quickly and saturate efficiently, with minimal trajectory spread. Coolant behavior shows tight regulation, minimal excursions, and no signs of instability, even during transient phases.

Compared to the baseline PI rollouts, BC policy, and BC+ corrected actions, IQL clearly dominates in both control quality and consistency. This result validates IQL as a strong candidate for real-world deployment, showcasing its ability to extract high-quality policies from offline data without requiring online correction or cost shaping. It serves as a practical gold standard for offline RL in process control settings where safety, sample efficiency, and asymptotic performance are all critical.

\subsubsection{IQL with online gradient-based action corrections}

\begin{figure}[htb]
	\centering
	\includegraphics[width=0.7\textwidth]{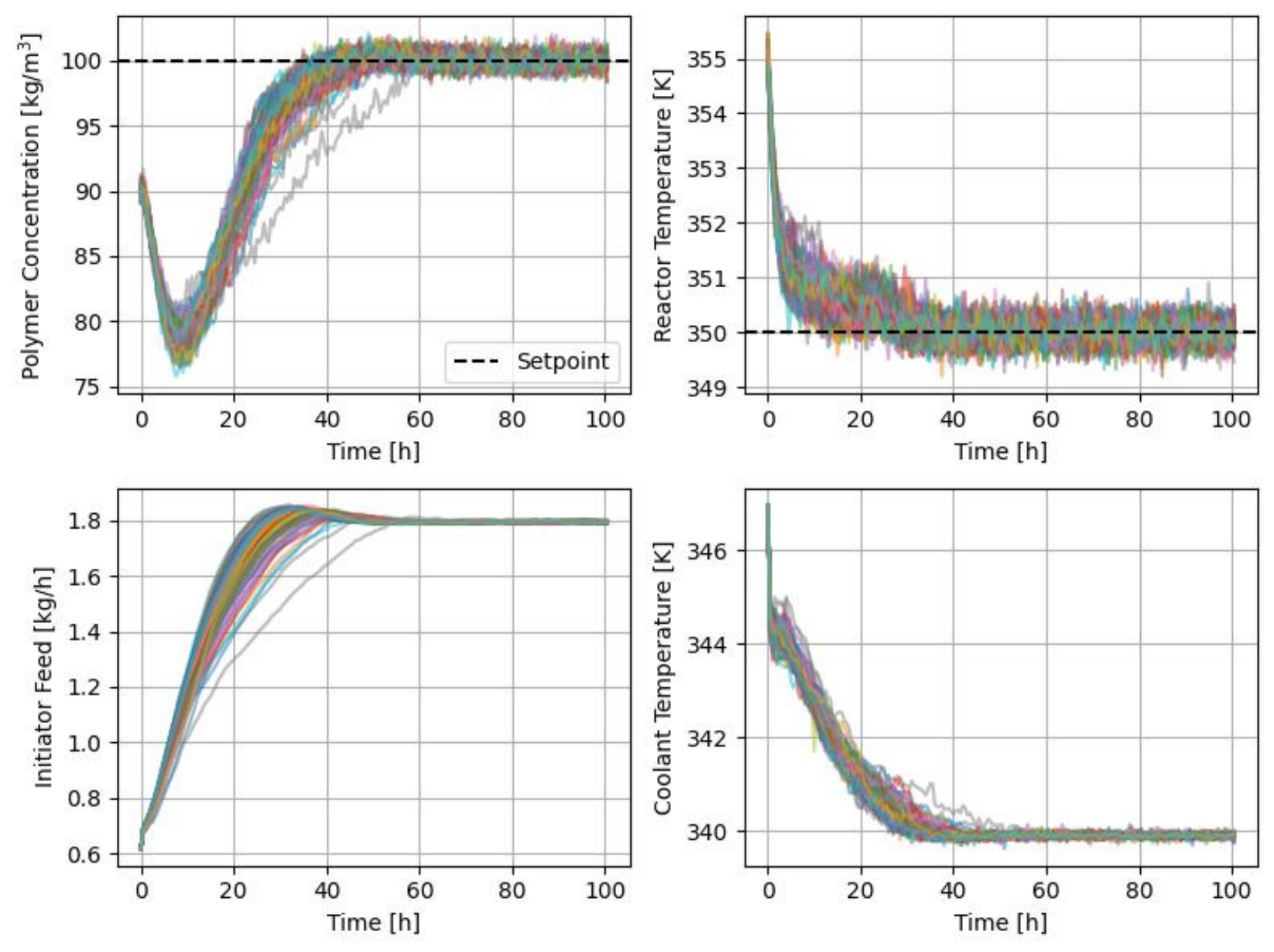}
	\caption{IQL agent with online gradient-based action corrections on the grade up scenario.}
	\label{fig:iqlp}
\end{figure}

Figure \ref{fig:iqlp} shows the results of augmenting the IQL policy with online gradient-based action corrections using the PICNN cost model. As seen, the performance remains largely on par with the uncorrected IQL controller, which already served as a strong baseline. However, subtle but meaningful refinements are apparent.

The most notable improvement lies in the polymer concentration trajectories (top-left). While IQL already exhibited excellent setpoint tracking, a small subset of trajectories previously lagged slightly in convergence. With the addition of PICNN-based corrections, these sluggish responses are effectively eliminated. The result is an even tighter clustering of trajectories around the setpoint, further reducing variance and improving worst-case performance without compromising stability.

In the reactor temperature (top-right), initiator feed (bottom-left), and coolant temperature (bottom-right) plots, the differences compared to standard IQL are minimal. This is expected as IQL already achieves near-optimal actuation patterns in these variables. Importantly, the cost-based corrections do not degrade performance or introduce instability, demonstrating the compatibility of PICNN refinements with high-quality base policies.

Overall, while there is little headroom to improve upon IQL's already strong performance, applying online corrections via a convex cost model provides an additional layer of robustness and fine-tuning. It acts as a safeguard against occasional suboptimal actions and offers a principled mechanism for enhancing policy behavior in edge cases, particularly in safety-critical or highly sensitive control loops like polymer concentration.

\subsubsection{Grade up scenario summary}

\begin{figure}[htb]
	\centering
	\includegraphics[width=0.8\textwidth]{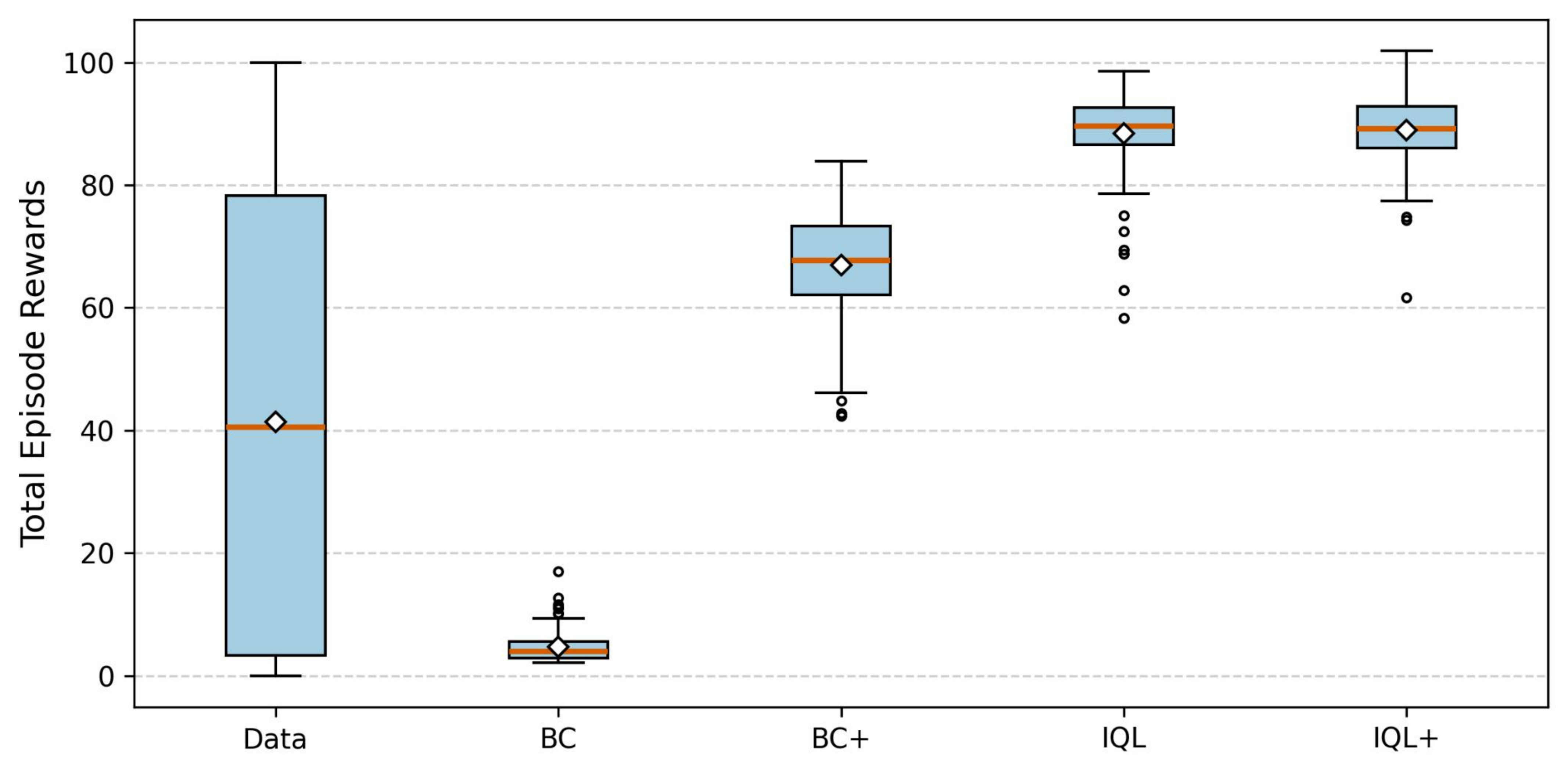}
	\caption{Total episode rewards for rollouts on the polymerisation CSTR grade up scenario. The boxplots show the distribution of rewards achieved by the PI controllers (Data), BC, IQL agents, and BC and IQL with online gradient-based action corrections (BC$+$ and IQL$+$) across 100 episodes. The median is shown as a central line, the mean is shown by a diamond marker, the box represents the IQR of the data, and the whiskers extend to the farthest data point lying within 1.5x the IQR from the box. Outliers beyond the end of the whiskers are shown as individual circle markers.}
	\label{fig:boxplots}
\end{figure}

Figure \ref{fig:boxplots} summarises the total episode reward distributions across all controllers, normalised such that the worst-performing trajectory in the PI-controlled dataset scores 0 and the best scores 100. This provides a consistent benchmark for evaluating learning-based controllers relative to the diversity of behaviours present in the offline data.

The baseline dataset (Data) shows a wide spread in performance, reflecting the highly variable quality of the random PI rollouts. The IQR spans nearly the full reward spectrum, with many low-quality trajectories pulling down the median score to around 40. This highlights the challenging nature of learning from such noisy, heterogeneous data.

BC performs poorly in this metric, with both mean and median rewards significantly below those of the dataset. The narrow spread and dense cluster of low scores indicate that BC frequently reproduces the conservative, sluggish behaviors present in the lower end of the data distribution as already seen in the polymer concentration responses (Figure \ref{fig:bc}).

Once the BC policy is augmented with PICNN-based gradient corrections (BC+), there is a dramatic improvement. The median reward jumps substantially, approaching the upper quartile of the original dataset. The variance is also reduced relative to the data, and notably, the worst-case performance improves by a large margin. This demonstrates that cost model corrections are highly effective in recovering much of the lost performance from a weak prior, transforming a non-viable offline RL agent into one that can compete in the upper half of the training data distribution.

IQL already achieves high reward scores with low variance, confirming its strong and consistent policy performance observed earlier (Figure \ref{fig:iql}). The range of episode rewards is tightly concentrated near the upper bound, indicating both efficiency and robustness across runs.

Applying online corrections to IQL (IQL+) provides an additional lift in the reward distribution. While the median remains nearly unchanged, both the minimum and maximum scores improve. This suggests that PICNN corrections can still add measurable benefit even on top of a strong offline RL policy, particularly by mitigating rare but costly suboptimal decisions. The PICNN-based action corrections significantly enhance BC policies and provide measurable robustness gains even for strong RL agents like IQL.

\clearpage

\subsection{Other scenarios}
Table \ref{tab:results} summarises the mean total rewards achieved by the different agents across all 3 polymerisation CSTR scenarios: startup, grade-down, and grade-up. All results are normalised relative to the PI-generated datasets for each scenario, with 0 corresponding to the worst and 100 to the best PI trajectory. This allows for meaningful comparisons of control quality across different operating regimes.

\begin{table}[htb]
\centering
\caption{Summary of mean total rewards (normalised on the PI-generated datasets) from the polymerisation CSTR control scenarios.}
\label{tab:results}
\begin{tabular}{lllll}
\hline
	& \textbf{Startup} & \textbf{Grade Down} & \textbf{Grade Up} & \textbf{Mean} \\
\hline
Data & 39.4 & 52.6 & 41.4 & 44.5 \\
BC & 41.2 & 72.6 & 4.70 & 39.5 \\
BC+ & 66.2 & 80.8 & 66.9 & 71.3 \\
IQL & \textbf{98.6} & 89.0 & 88.4 & 92.0 \\
IQL+ & 96.7 & \textbf{90.4} & \textbf{88.9} & \textbf{92.3} \\
\hline
\end{tabular}
\end{table}

In the startup scenario, the IQL agent achieved the highest mean total reward (98.6), demonstrating exceptional performance. Interestingly, the IQL+ agent (which includes online gradient-based action corrections using a PICNN cost model) achieved a slightly lower mean score (96.7). However, the minimum total reward for IQL+ was notably higher (81.1 compared to 72.7), indicating improved consistency and robustness. This suggests that while online corrections may not always raise average performance when starting from a strong policy, they can reduce worst-case deviations, which is particularly valuable in safety-critical applications.

In the grade-down scenario, both IQL and IQL+ again performed strongly, with IQL+ achieving the highest mean reward (90.4), slightly outperforming IQL (89.0). The benefits of online correction are again seen in the tightened lower end of the performance distribution. Meanwhile, BC achieved surprisingly strong results in this scenario (72.6), likely because the dynamics during grade-down transitions are more forgiving and less sensitive to timing errors. Still, BC+ provided a clear boost over BC (80.8 compared to 72.6), confirming that action refinement via gradient descent on a learned PICNN cost model improves performance even when the base behaviour is relatively effective.

In the grade-up scenario, the performance differences were most pronounced. The BC agent performed poorly (4.7 mean reward), reflecting its inability to handle the faster dynamics and more aggressive control required. However, applying PICNN-based corrections (BC+) dramatically improved its mean performance to 66.9 (more than a 14$\times$ increase) once again highlighting the utility of PICNN cost-driven refinement. IQL and IQL+ both achieved strong results here (88.4 and 88.9 respectively), with IQL+ slightly outperforming on average.

Across all scenarios, the IQL+ agent attained the highest mean total reward overall (92.3), followed closely by IQL (92.0). These results support the view that offline RL is capable of producing high-performing and generalisable policies from diverse datasets. Moreover, the marginal yet consistent gains provided by IQL+ show that combining these policies with structure-preserving cost models enables more robust deployment, especially under dynamic or uncertain conditions.

In contrast, the BC agent alone underperformed (overall mean 39.5), frequently replicating suboptimal behavior from the dataset. Yet, when augmented with PICNN gradient corrections (BC+), its mean reward increased dramatically to 71.3 reinforcing that PICNN cost model action corrections can rescue otherwise suboptimal and potentially unsafe offline policies.

\section{Conclusion} \label{sec:conclusion}

This work demonstrates the feasibility and promise of applying offline RL to the safe and efficient control of chemical process systems, using a polymerisation CSTR as a representative case study. We developed a realistic, Gymnasium-compatible simulation environment and generated diverse offline datasets using PI controllers, establishing a robust testbed for evaluating offline RL algorithms under industrially relevant conditions. BC served as a strong imitation baseline but was limited by its dependence on the quality and coverage of the offline data. In contrast, IQL leveraged value-based learning and advantage-weighted regression to outperform BC in terms of robustness and generalisation.

To address the safety and stability challenges inherent in deploying offline RL policies, we introduced a practical and principled mechanism for real-time action correction using gradient-based updates over a learned cost model. By structuring the cost function as a PICNN, we ensured convexity in the action space while retaining rich expressivity over system states. This architectural choice enables stable, interpretable, and efficient action refinement at deployment time without requiring additional environment interaction or explicit constraint programming.

Empirical results show that enforcing convexity in the cost model leads to smoother optimisation landscapes, more reliable corrective updates, and improved constraint satisfaction, all while maintaining task performance. These results highlight the potential of offline RL as a data-efficient alternative to traditional control strategies, particularly in complex transition scenarios where classical methods like PI control may require extensive tuning or fail to generalise.

Looking forward, this work opens several promising research directions. Future efforts will focus on extending our method to multi-step or trajectory-based safety objectives, incorporating uncertainty estimation to improve robustness, and generalising convexity-aware correction to broader classes of value or reward models. Additionally, we are interested in exploring hybrid control architectures that combine the interpretability and reliability of classical control with the flexibility and adaptability of learning-based approaches. Integrating domain knowledge, safety constraints, and uncertainty quantification will be key to advancing the safe deployment of RL in real-world industrial settings.

\section*{Acknowledgements}

% Acknowledgements have been ommited for review purposes.
The authors would like to acknowledge the financial support from Shell.

% references
\bibliographystyle{IEEEtran}
\bibliography{references}

\clearpage

\appendix

\section{Startup scenario results} \label{sec:su_results}

% data_startup
\begin{figure}[htb]
    \centering
    \includegraphics[width=0.7\textwidth]{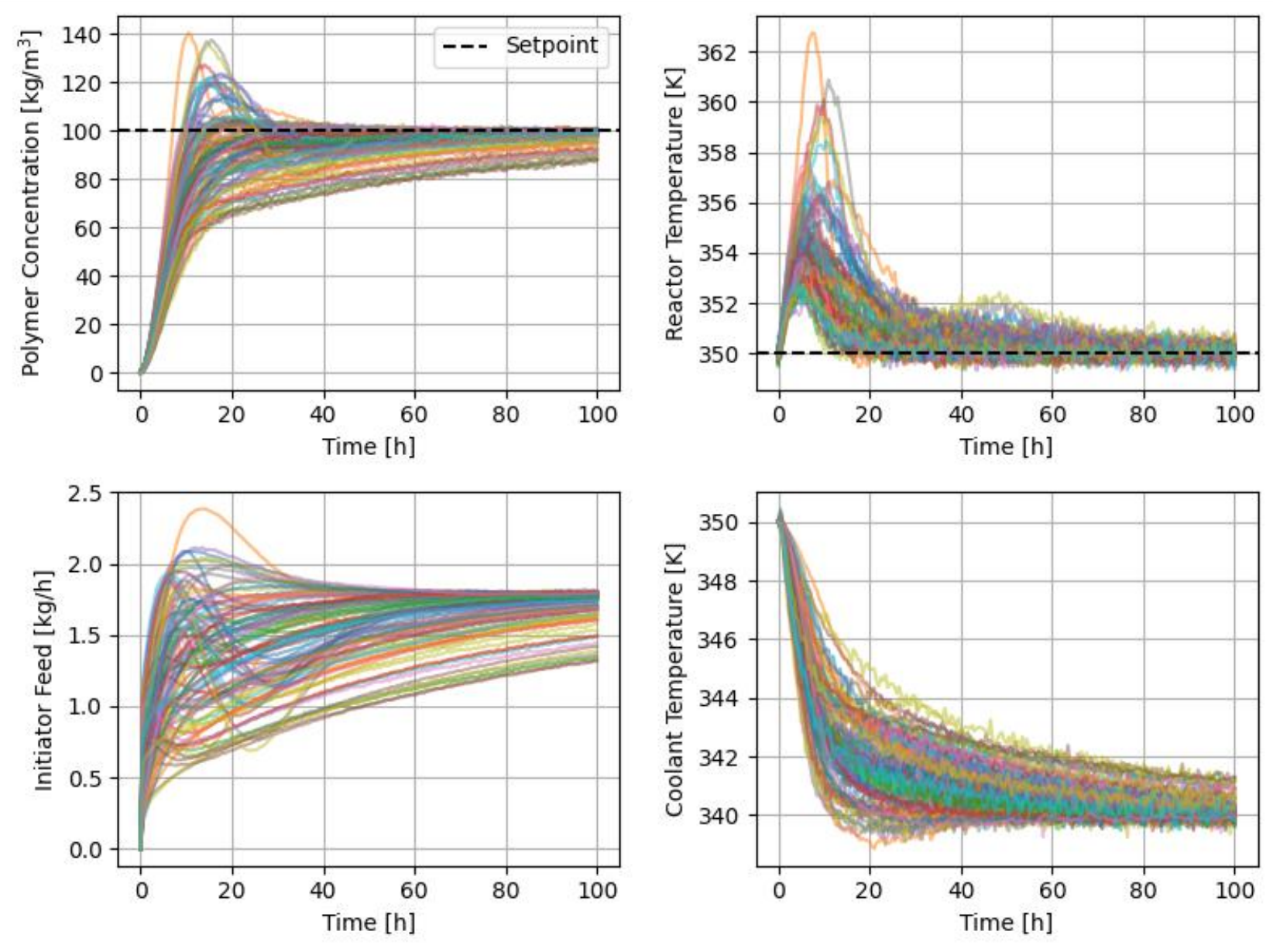}
    \caption{100 PI controlled episodes from the polymerisation CSTR startup scenario.}
    \label{fig:data_startup}
\end{figure}

% bc_startup
\begin{figure}[htb]
    \centering
    \includegraphics[width=0.7\textwidth]{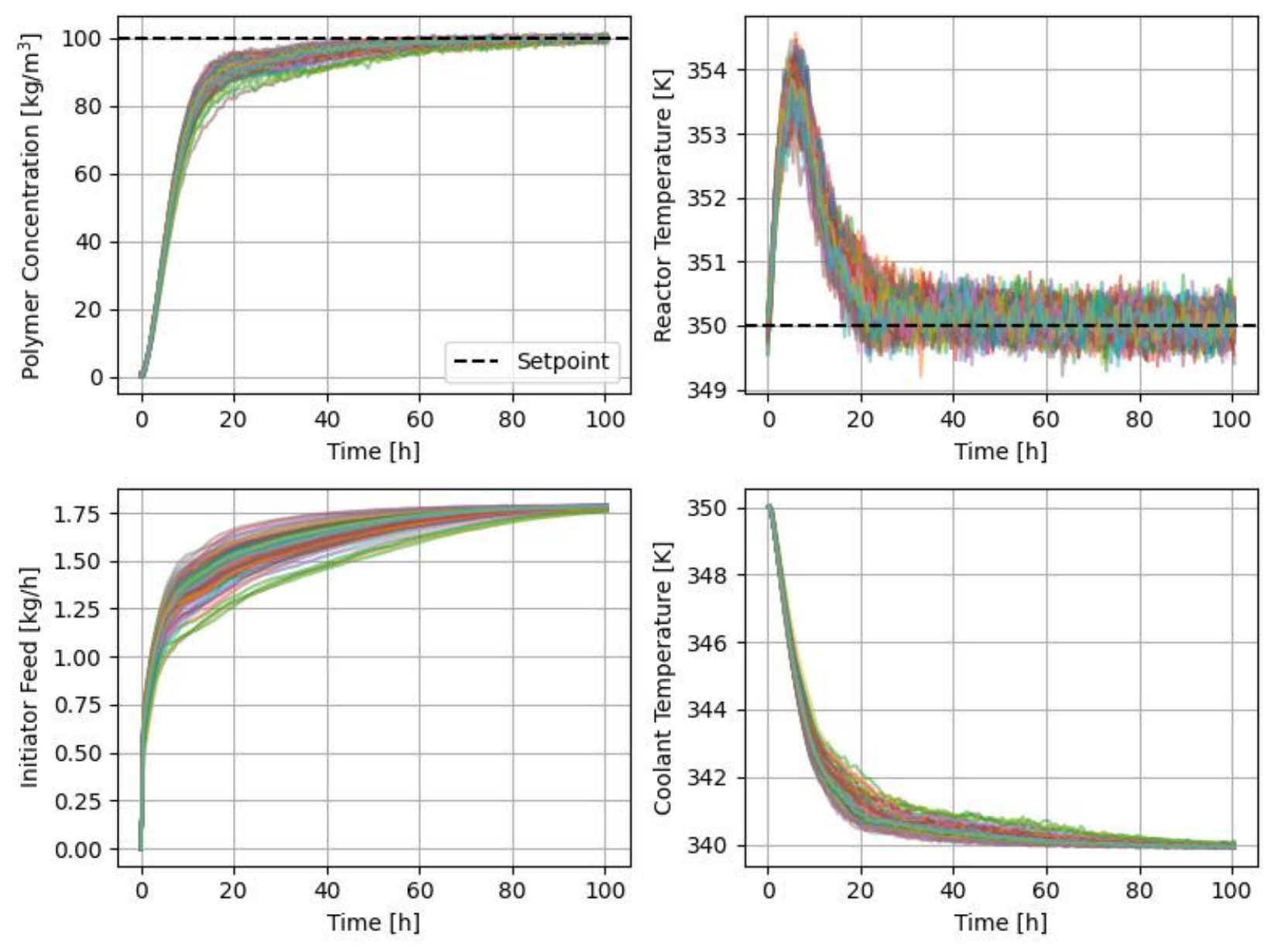}
    \caption{100 BC episodes for the startup scenario.}
    \label{fig:bc_startup}
\end{figure}

% iql_startup
\begin{figure}[htb]
    \centering
    \includegraphics[width=0.7\textwidth]{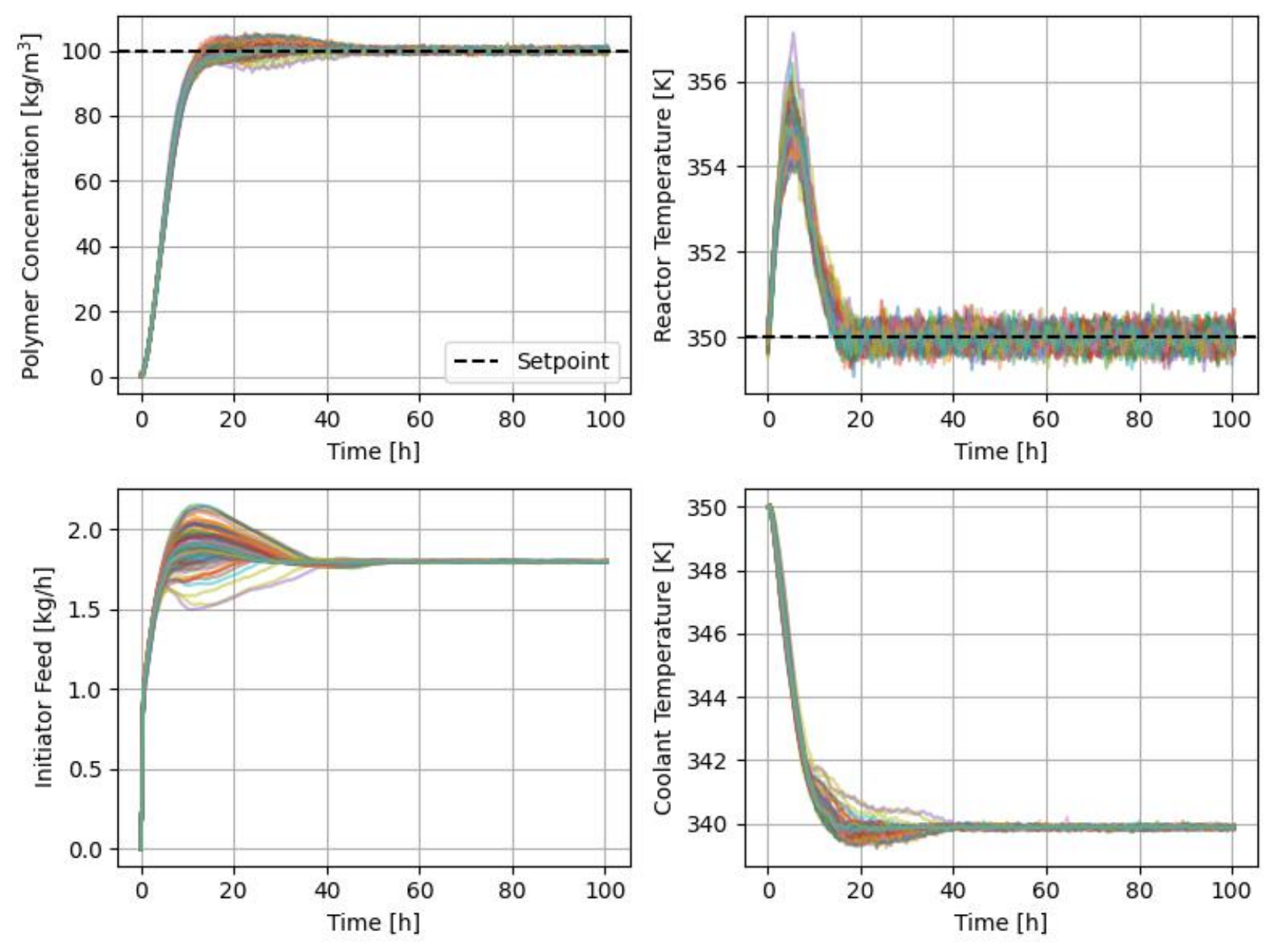}
    \caption{100 IQL episodes for the startup scenario.}
    \label{fig:iql_startup}
\end{figure}

\begin{figure}[htb]
    \centering
    \includegraphics[width=0.7\textwidth]{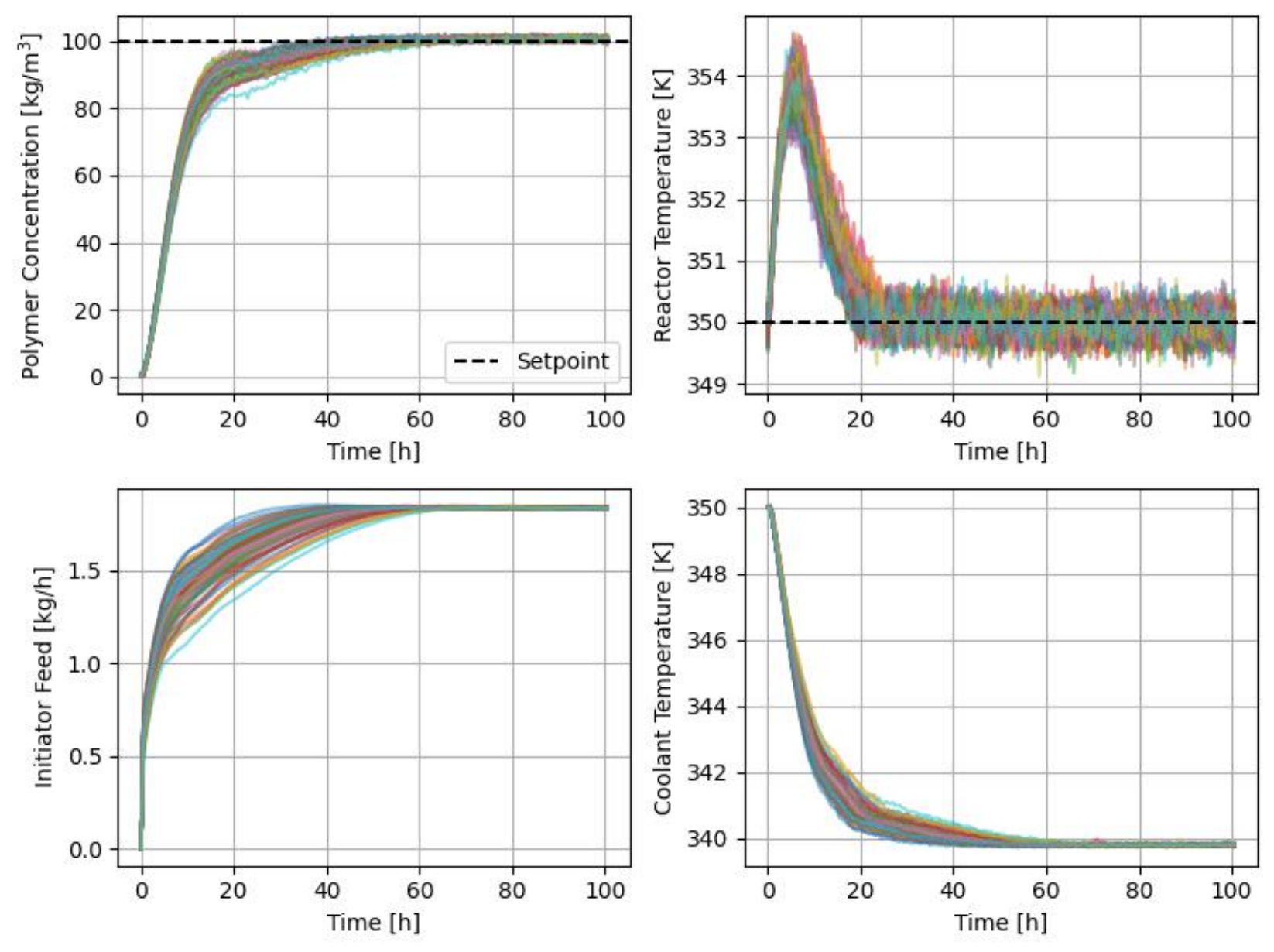}
    \caption{100 episodes for the startup scenario controlled by the BC agent with online gradient correction.}
    \label{fig:bcp_startup}
\end{figure}

\begin{figure}[htb]
    \centering
    \includegraphics[width=0.7\textwidth]{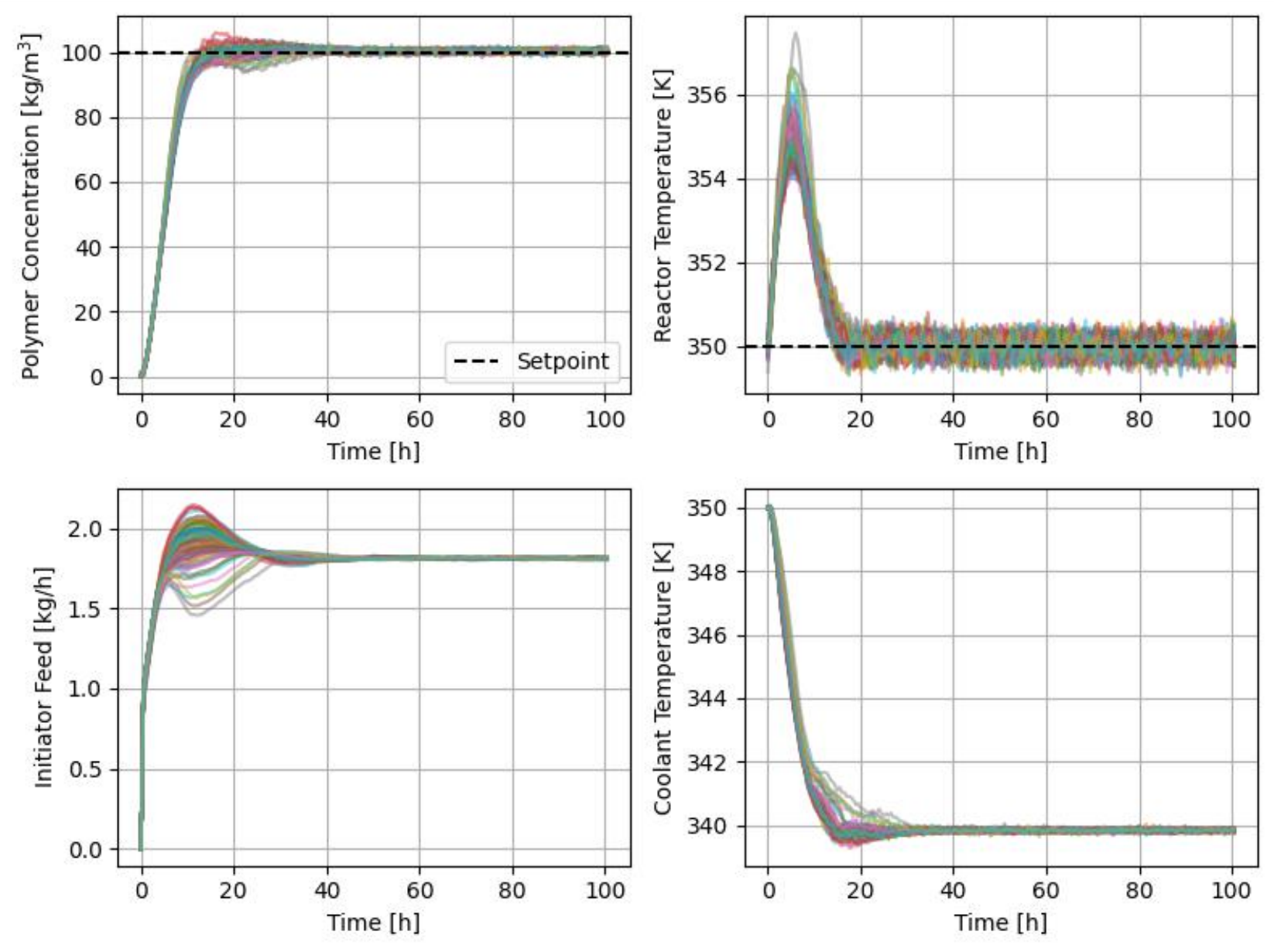}
    \caption{100 episodes for the startup scenario controlled by the IQL agent with online gradient correction.}
    \label{fig:iqlp_startup}
\end{figure}

\begin{figure}[htb]
    \centering
    \includegraphics[width=0.7\textwidth]{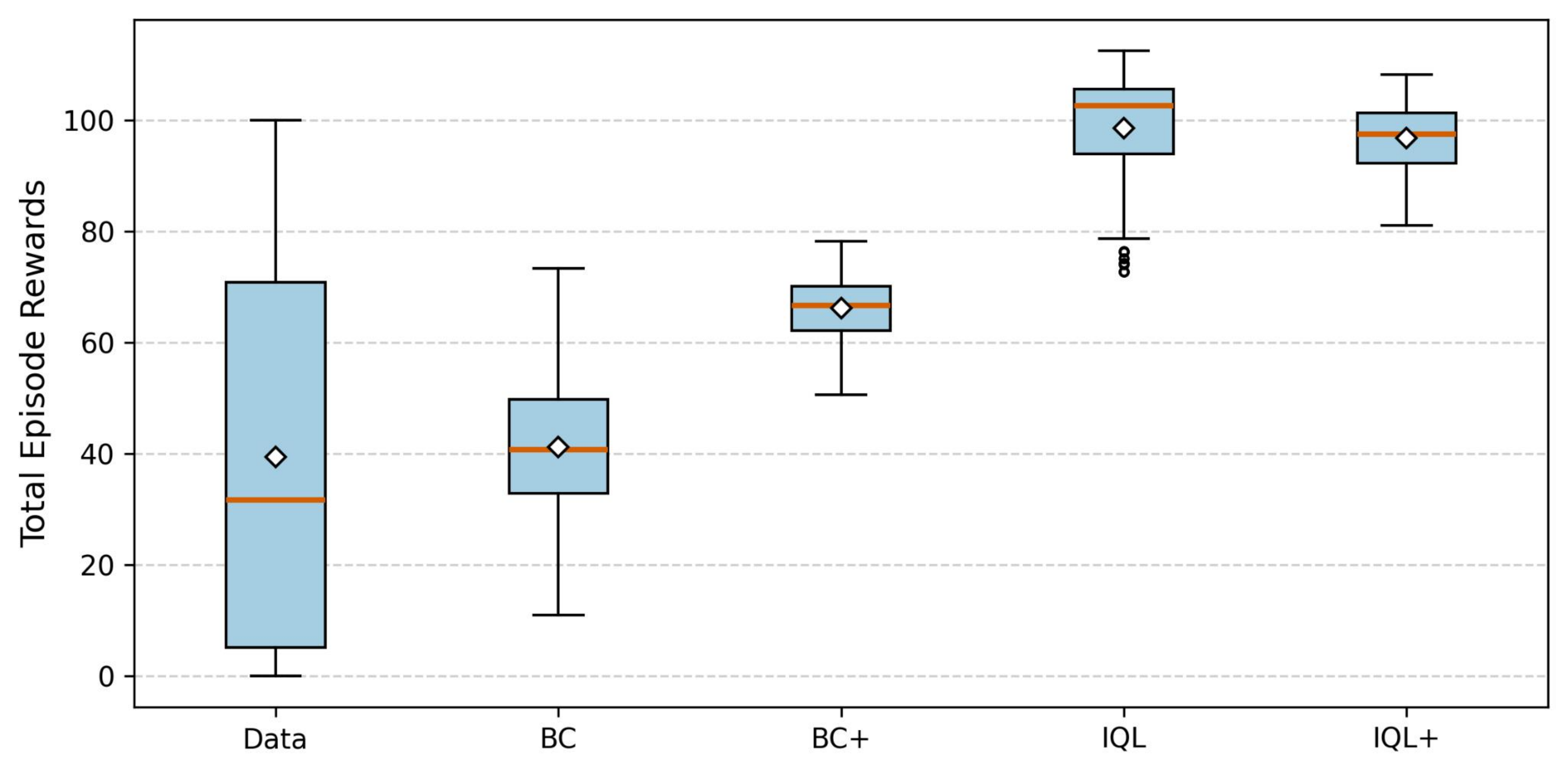}
    \caption{Boxplot of rewards for the startup scenario.}
    \label{fig:boxplot_startup}
\end{figure}

\clearpage
\section{Grade down scenario results} \label{sec:gd_results}

% data_grade_down
\begin{figure}[htb]
    \centering
    \includegraphics[width=0.7\textwidth]{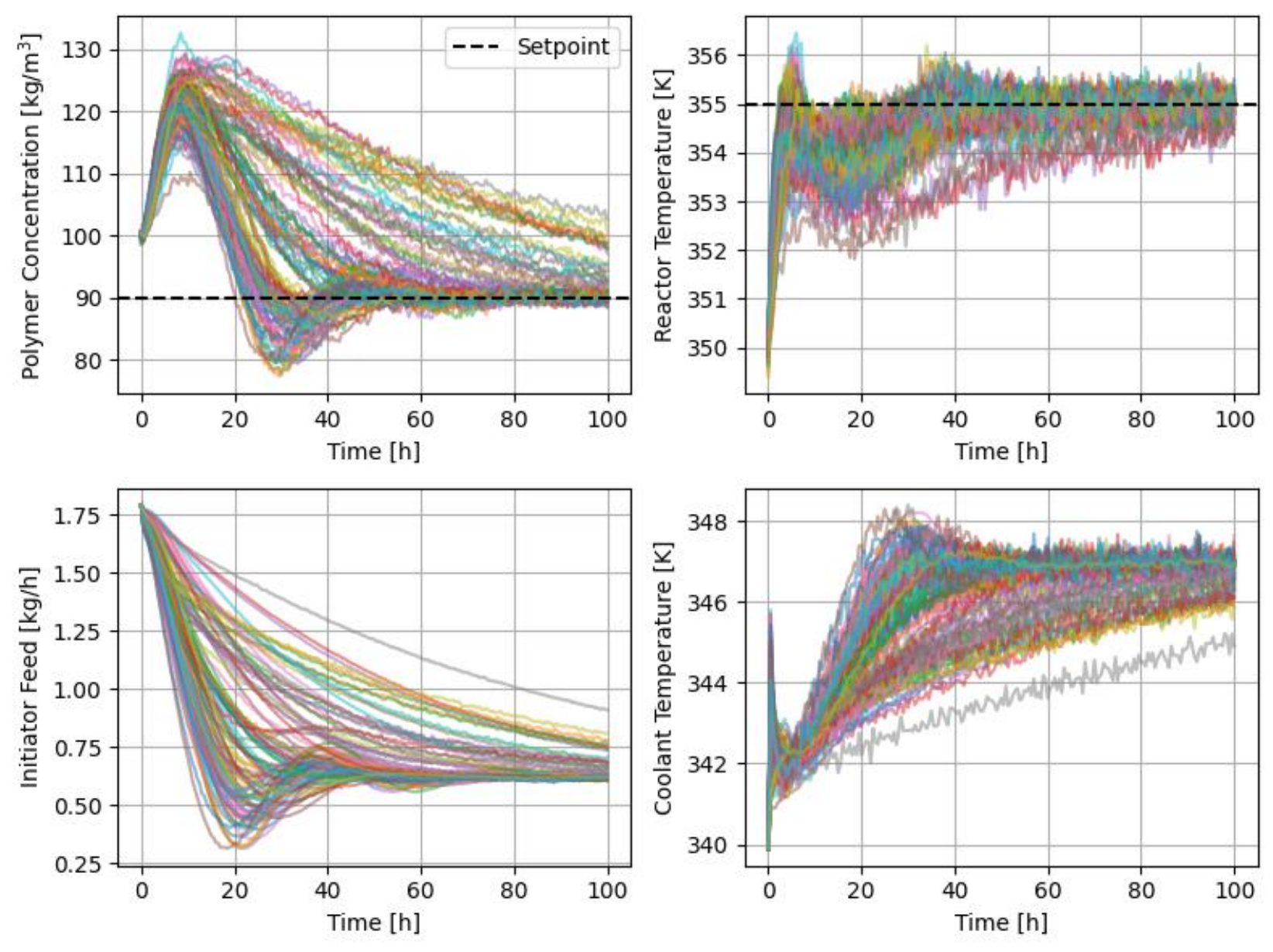}
    \caption{100 PI controlled episodes from the polymerisation CSTR grade change down scenario.}
    \label{fig:data_grade_change_down}
\end{figure}

% bc_grade_down
\begin{figure}[htb]
    \centering
    \includegraphics[width=0.7\textwidth]{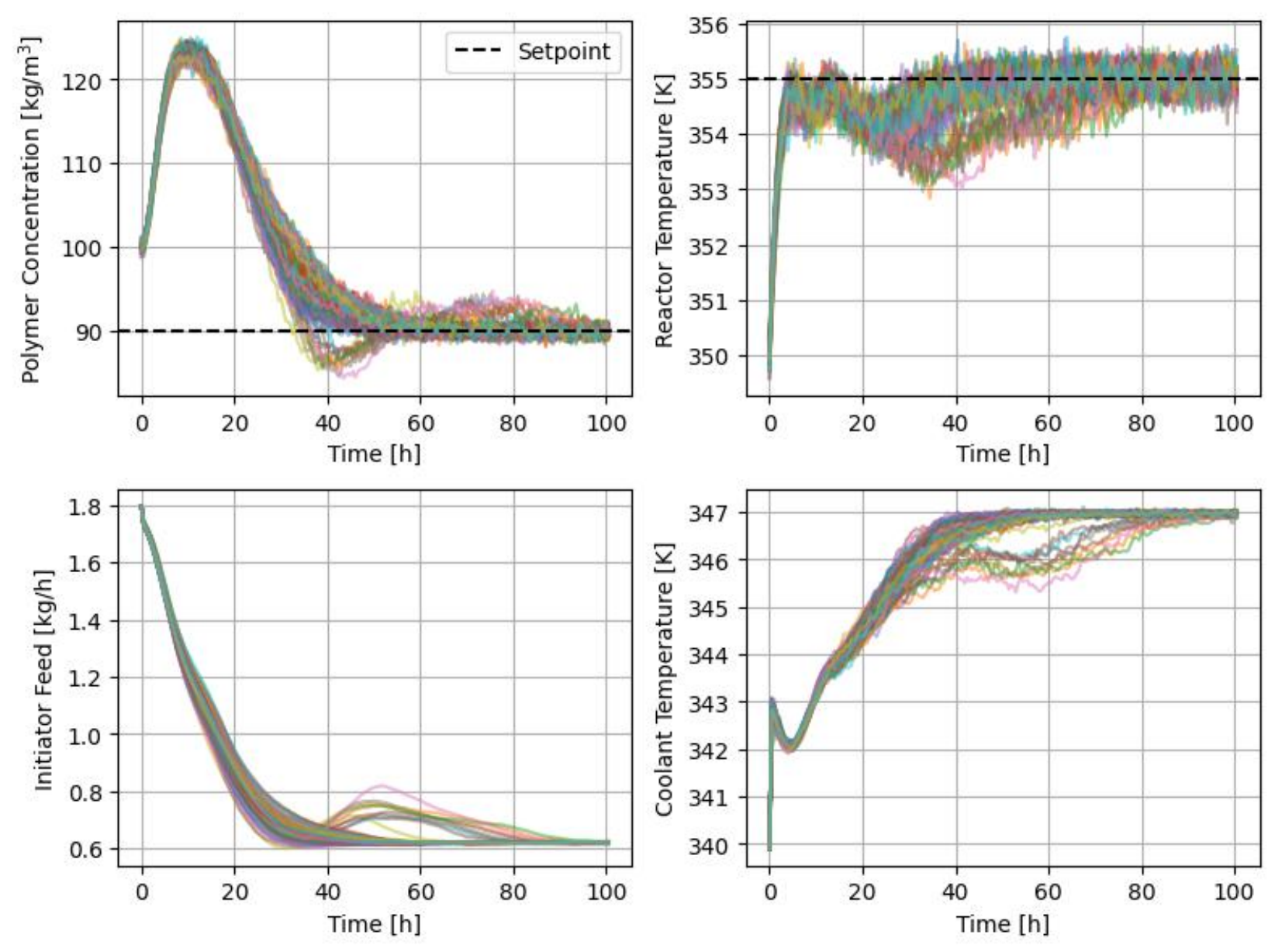}
    \caption{100 BC episodes for the grade change down scenario.}
    \label{fig:bc_grade_down}
\end{figure}

% iql_grade_down
\begin{figure}[htb]
    \centering
    \includegraphics[width=0.7\textwidth]{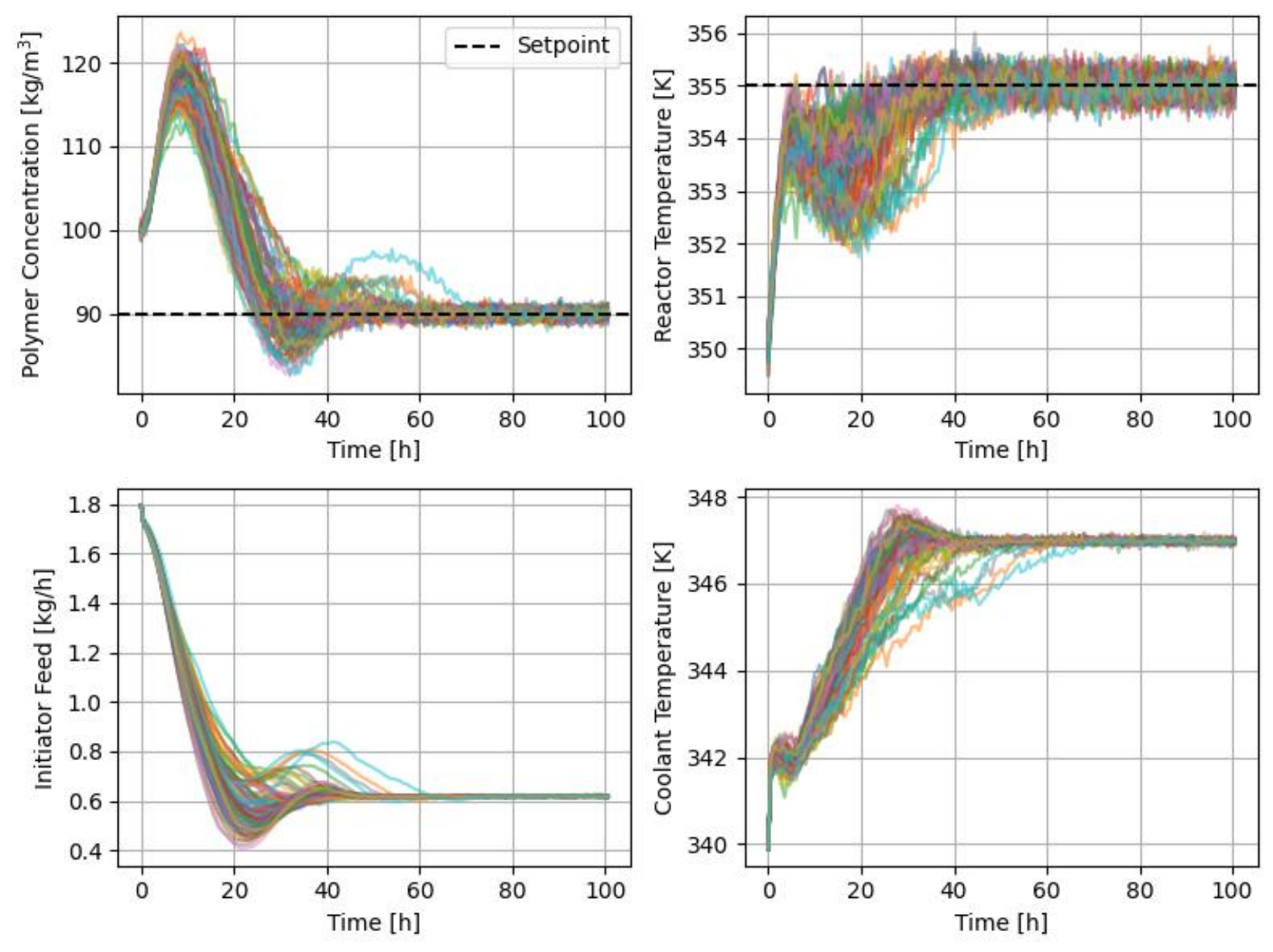}
    \caption{100 IQL episodes for the grade change down scenario.}
    \label{fig:iql_grade_down}
\end{figure}

% bcp_grade_down
\begin{figure}[htb]
    \centering
    \includegraphics[width=0.7\textwidth]{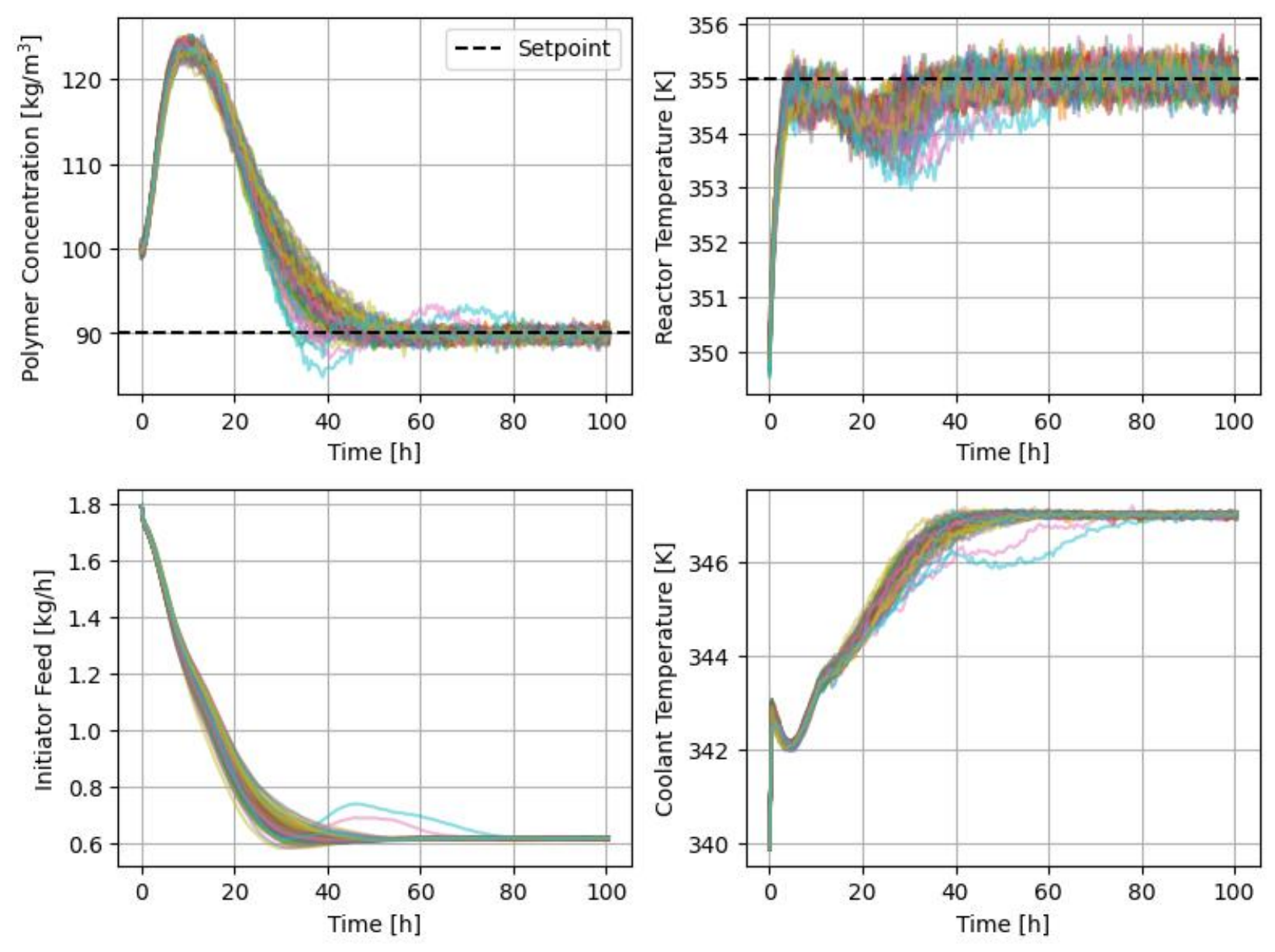}
    \caption{100 episodes for the grade change down scenario controlled by the BC agent with online gradient correction.}
    \label{fig:bcp_grade_down}
\end{figure}

% iqlp_grade_down
\begin{figure}[htb]
    \centering
    \includegraphics[width=0.7\textwidth]{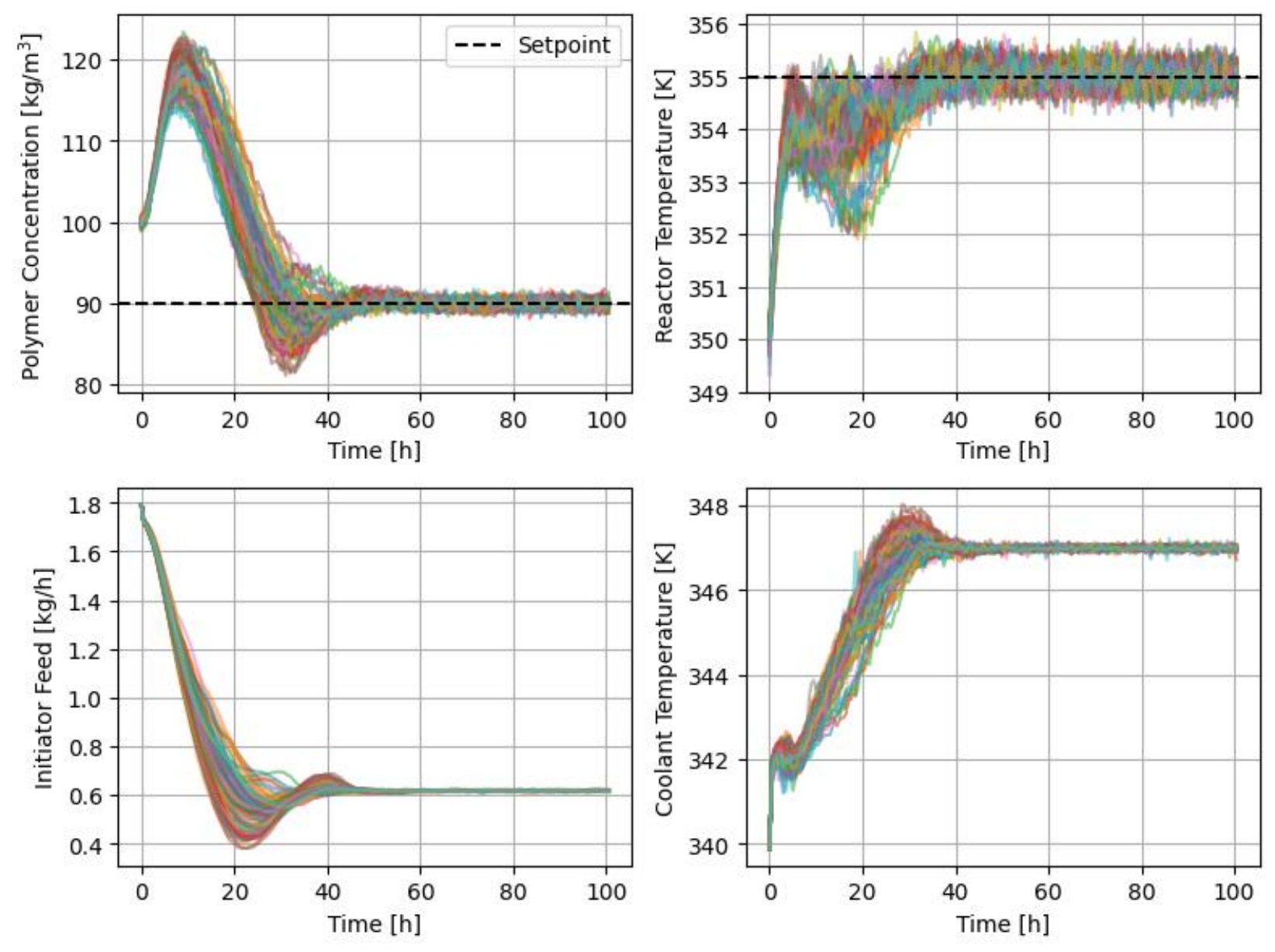}
    \caption{100 episodes for the grade change down scenario controlled by the IQL agent with online gradient correction.}
    \label{fig:iqlp_grade_down}
\end{figure}

\begin{figure}[htb]
    \centering
    \includegraphics[width=0.7\textwidth]{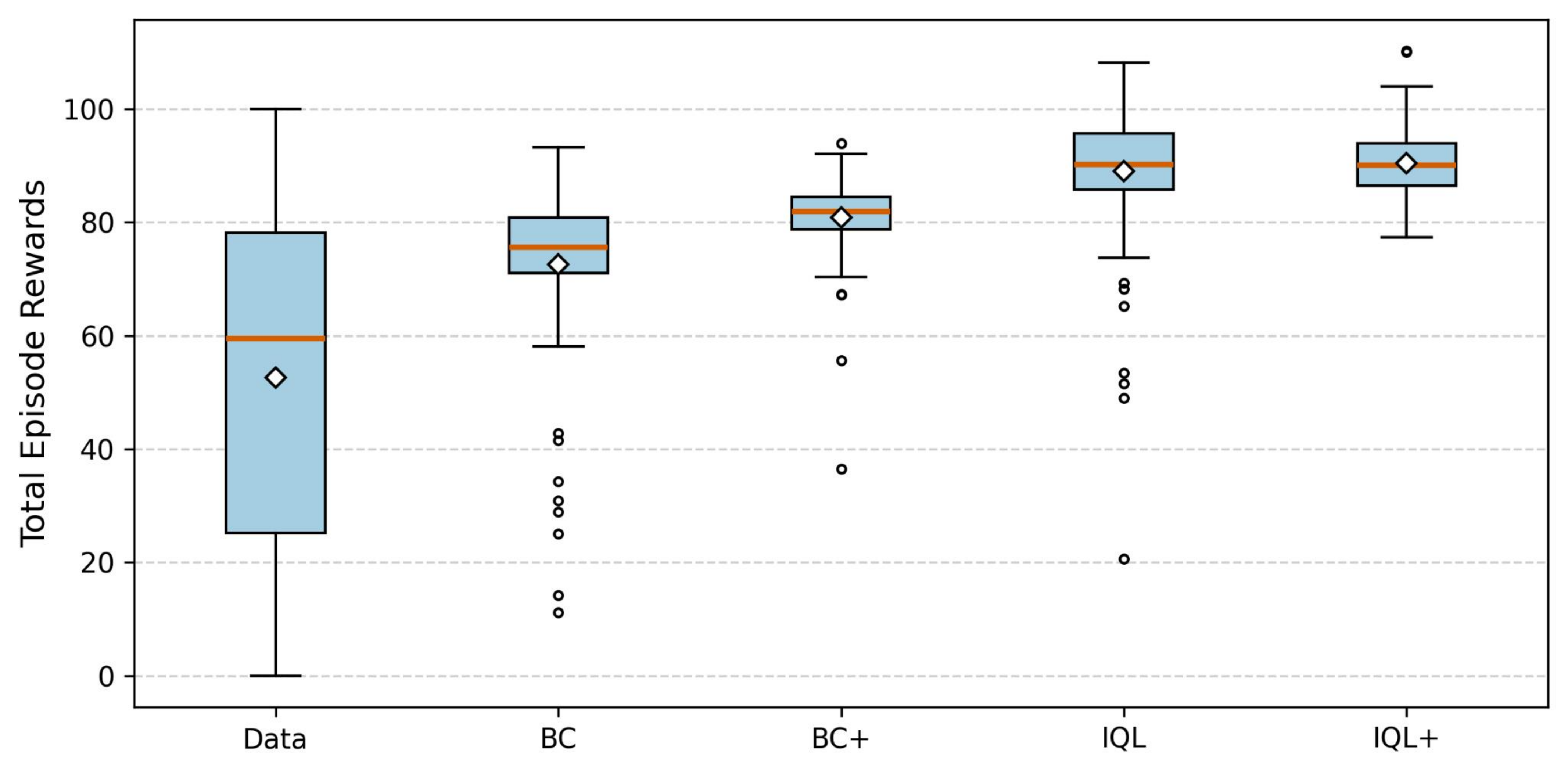}
    \caption{Boxplot of rewards for the grade change down scenario.}
    \label{fig:boxplot_grade_down}
\end{figure}

\end{document}